\documentclass{article}
\usepackage{amssymb}
\usepackage{amsfonts}
\usepackage{amsmath}
\usepackage{epsfig}
\usepackage{graphicx}

\begin{document}

\title{Characterizing correlations and synchronization in collective dynamics%
}
\author{\textbf{Carlos Aguirre} \thanks{%
e-mail: carlos.aguirre@uam.es} \\
{\small GNB, Escuela Polit\'{e}cnica Superior, }\\
{\small \ Universidad Autonoma de Madrid, Campus de Cantoblanco, }\\
{\small \ Ctra de Colmenar Km 16, 28049 Madrid, Spain} \and \textbf{R.
Vilela Mendes} \thanks{%
e-mail: rvilela.mendes@gmail.com} \thanks{%
Corresponding author} \\
{\small CMAF, Universidade de Lisboa,}\\
{\small Faculdade de Ci\^{e}ncias C6, 1749-016 Lisboa, Portugal}\\
{\small IPFN, Instituto Superior T\'{e}cnico, }\\
{\small Av. Rovisco Pais 1, 1049-001 Lisboa}}
\date{ }
\maketitle

\begin{abstract}
Synchronization, that occurs both for non-chaotic and chaotic systems, is a
striking phenomenon with many practical implications in natural phenomena.
However, even before synchronization, strong correlations occur in the
collective dynamics of complex systems. To characterize their nature is
essential for the understanding of phenomena in physical and social
sciences. The emergence of strong correlations before synchronization is
illustrated in a few piecewise linear models. They are shown to be
associated to the behavior of ergodic parameters which may be exactly
computed in some models. The models are also used as a testing ground to
find general methods to characterize and parametrize the correlated nature
of collective dynamics.
\end{abstract}


\section{Introduction}

Synchronization in complex systems \cite{Pikov1} \cite{Pikov2} \cite{Wu} 
\cite{Kocarev} \cite{Luo} \cite{Aguirre} \cite{Koronov} is a striking
cooperative phenomena in Nature that has been shown to be of fundamental
importance in fields as diverse as the operation of heart pacemaker cells 
\cite{Glass} \cite{Hayes}, circadian cycles \cite{circadian}, epileptic
seizures \cite{Schnitzler} \cite{Ortega}, schizophrenia disorders \cite%
{Ioannides} \cite{Sawa}, neuronal firing \cite{Roelfsema} \cite{Lachaux} 
\cite{Long}, animal behavior \cite{Buck}, social fads, the integration of
cognitive tasks \cite{Varela1} \cite{Varela2} \cite{Engel},
synchronization-based computation \cite{computation} and even quantum
systems \cite{quantum}.

Many natural systems can be described as networks of oscillators coupled to
each other. Coupled oscillators may display synchronized behavior, i.e.
follow a common dynamical evolution. Synchronization properties are
dependent on the coupling pattern among the oscillators, represented as an
interaction network \cite{synchro1} \cite{synchro2} \cite{synchro3} \cite%
{synchro4} \cite{synchro5} \cite{synchro6} \cite{synchro7} \cite{layer1} \ 
\cite{layer2}. The central question concerns the emergence of coherent
behavior: synchronization or other types of correlation. This occurs both
for systems with regular behavior as well as for systems which have chaotic
dynamics (lasers, neural networks, physiological processes, etc.). Chaotic
systems are characterized by a strong sensitivity to initial conditions, and
two identical uncoupled chaotic systems will become uncorrelated at large
times even if they start from very similar (but not identical) states.
Nevertheless, the coupling of such systems can make them follow the same
chaotic trajectories \cite{Fujisaka} \cite{Pecora1} \cite{Vilela1} \cite%
{Bocca} \cite{Pecora2} \cite{mobile}.

The degree of synchronization is usually measured by a parameter related to
the coherence of the phases or by the entropy of the phases distribution 
\cite{Rodriguez}. Most of the work developed so far in this field has
emphasized criteria for synchronizability and the relation between network
structure and the emergence of synchronized behavior. Typically, the
emphasis has been on the distinction between synchronized and incoherent
behavior or on their coexistence as in the so-called chimera states \cite%
{chimera1} \cite{chimera2} \cite{chimera3} \cite{chimera4} \cite{chimera5} 
\cite{chimera6}. Some exploration, mostly numerical, has also been done on
partially synchronized states, clustering, dimensional reduction, etc \cite%
{cluster1} \cite{cluster2} \cite{cluster3} \cite{cluster4} \cite{cluster5} 
\cite{cluster6} \cite{WataStro} \cite{Antonsen}. However little has been
done on the way of developing effective tools to characterize, in a
quantitative manner, the striking correlation phenomena that may appear
before synchronization or even in the apparently incoherent phases of some
systems. That is the main purpose of this work.

As a first step we have started to identify a set of models which, while
representative of the actual situations found in the natural world, were
also sufficiently simple to provide an identification of the mechanisms
underlying the correlation and synchronization phenomena. This search led to
the choice of:

A) A deformed Kuramoto model,

B) A model of coupled oscillators with a triangle interaction,

C) An integrate-and-fire model.

The the first two models are representative of the stylized behavior found
in many collective systems in biology, population dynamics, socio-economic
phenomena, etc. In addition, the first model has in some limit a complete
characterization of the Lyapunov spectrum, which provides a good hint on the
relevance of the ergodic invariants to collective behavior. Finally the
third model relates to neuroscience.

Once these models studied, mostly by numerical simulation, we have tested
wheter simple correlation tools might be sufficient to unravel the
correlation behavior that arises before synchronization. Having these tests,
partly described in subsection 2.1, shown that the usual correlation
measures are not sufficient for our purposes, we set out to develop new
tools which are described in section 3 and tested on the models in section
4. Admittedly, the new tools are more complex than simple correlations.
Nevertheless they are not hard to program and even implement as an automatic
diagnostic.

\section{The models}

\subsection{A deformed Kuramoto model}

The main model used, in the past, for the study of synchronization phenomena
was the Kuramoto model \cite{Kuramoto}%
\begin{equation}
\frac{d\theta _{i}}{dt}=\omega _{i}+\frac{K}{N-1}\sum_{j=1}^{N}\sin \left(
\theta _{j}-\theta _{i}\right)  \label{1.1}
\end{equation}%
The analysis of the Kuramoto model has a long history, with a number of
important results obtained throughout the years \cite{Strogatz1} \cite%
{Acebron} but a full understanding of its dynamics is still lacking, and
most of the rigorous results are only strictly valid in the thermodynamic
limit.

\begin{figure}[htb]
\centering
\includegraphics[width=0.6\textwidth]{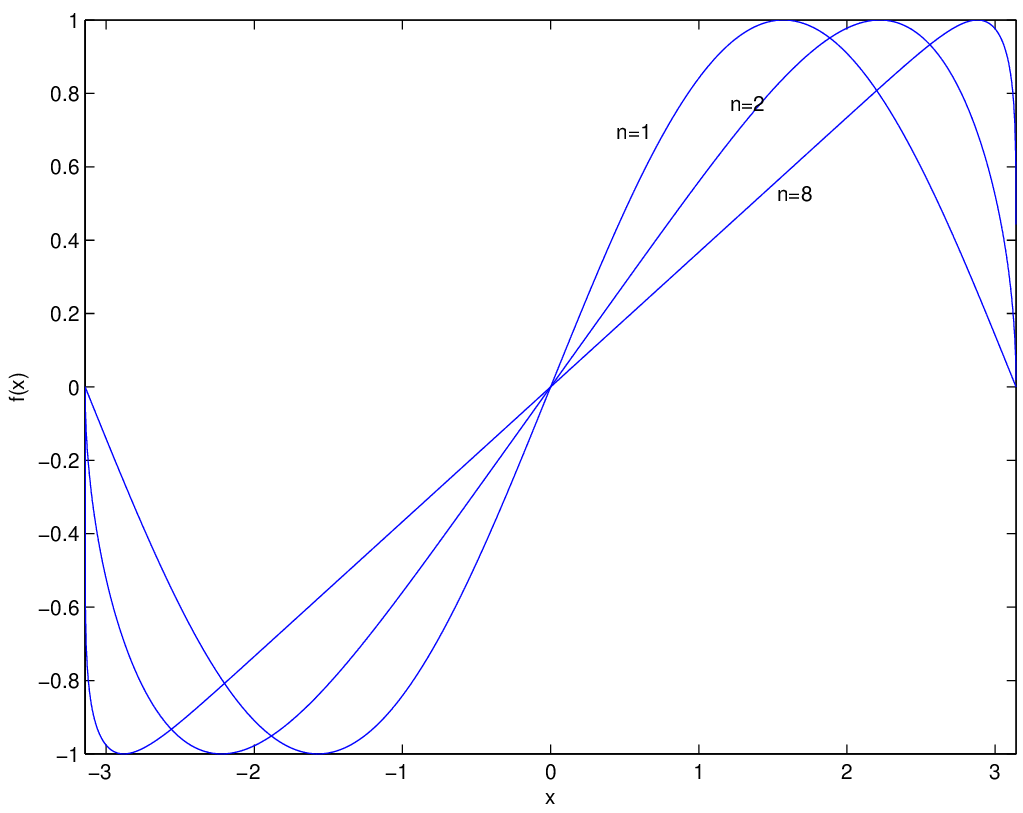}
\caption{The $f^{\left( n\right)}$ interaction function}
\label{f_x}
\end{figure}

Here we use a model of the same type. This model, first mentioned in \cite%
{Vilela3}, is%
\begin{equation}
x_{i}\left( t+1\right) =x_{i}\left( t\right) +\omega _{i}+\frac{K}{N-1}%
\sum_{j=1}^{N}\pi f^{(n)}\left( x_{j}-x_{i}\right) \hspace{1cm}(\mathnormal{%
mod}\pi )  \label{2.1}
\end{equation}%
with $x_{i}\in \lbrack -\pi ,\pi )$ and $f^{\left( n\right) }$ a deformed
version of the Kuramoto interaction%
\begin{equation}
f^{\left( n\right) }\left( x\right) =\text{sign}\left( x\right) \left( \sin
\left( \frac{\left\vert x\right\vert ^{n}}{\pi ^{n-1}}\right) \right) ^{1/n}
\label{2.2}
\end{equation}%
For $n=1$ $f^{\left( 1\right) }=\sin \left( x\right) $ and when $%
n\rightarrow \infty $ it becomes (Fig. \ref{f_x})

\begin{equation}
f^{(\infty )}\left( x_{j}-x_{i}\right) =\frac{1}{\pi }\left(
x_{j}-x_{i}\right) \hspace{1cm}\left( \text{mod }1\right)  \label{2.3}
\end{equation}%
For the numerical examples the $\omega _{i}^{\prime }s$ will follow a Cauchy
distribution%
\begin{equation}
p\left( \omega \right) =\frac{\gamma }{\pi \left[ \gamma ^{2}+\left( \omega
-\omega _{0}\right) ^{2}\right] }  \label{2.4}
\end{equation}%
The $f^{(\infty )}$ interaction will be used to derive parameters that
characterize the correlations that emerge before synchronization. However
they can also be easily computed in more general systems.

For coupled dynamical systems, an order parameter for synchronization is,
for example \cite{Restrepo}%
\begin{equation}
R_{n}\left( t\right) =\left\vert \sum_{m=1}^{N}A_{n,m}e^{ix_{m}\left(
t\right) }\right\vert  \label{2.5}
\end{equation}%
where $A$ is the adjacency matrix. For the fully coupled system we consider,
it is simply%
\begin{equation}
r\left( t\right) =\left\vert \frac{1}{N}\sum_{j=1}^{N}e^{ix_{j}\left(
t\right) }\right\vert  \label{2.6}
\end{equation}

\begin{figure}[htb]
\centering
\includegraphics[width=0.6\textwidth]{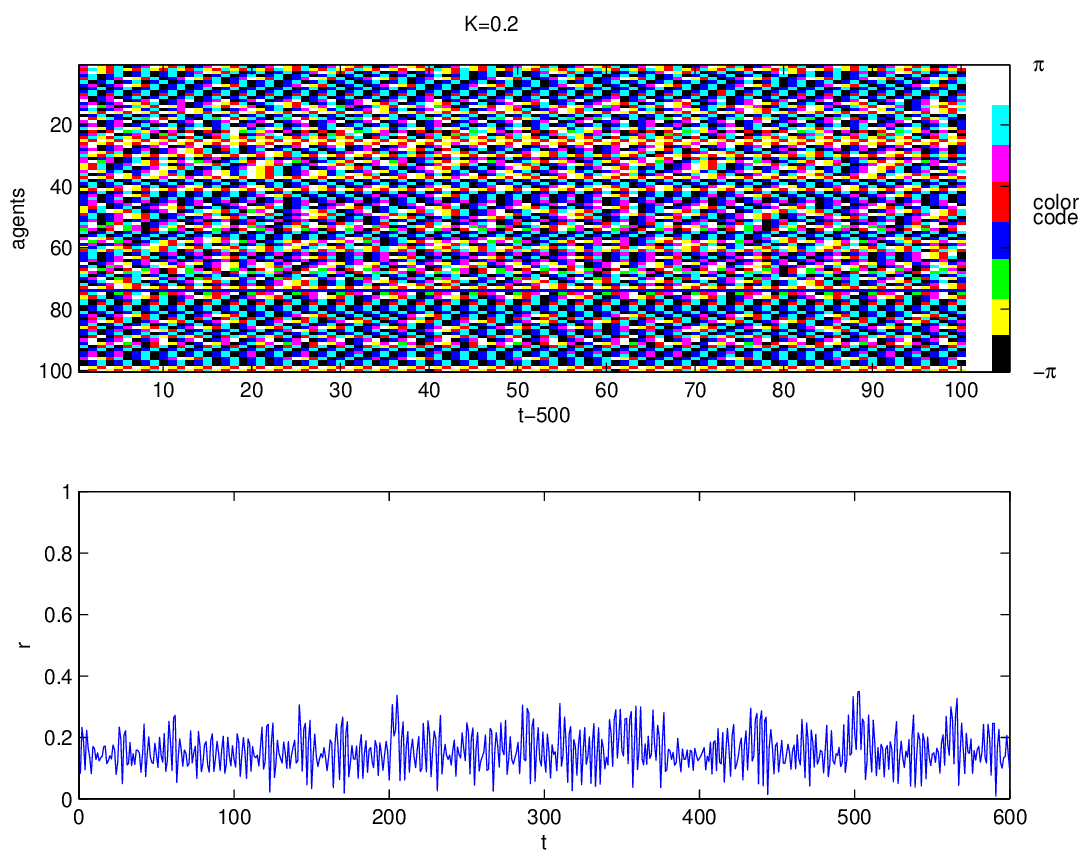}
\caption{Coordinates $x_{i}$ and order parameter $r\left( t\right) $ for $%
K=0.2$}
\label{sync_0_2}
\end{figure}

\begin{figure}[htb]
\centering
\includegraphics[width=0.6\textwidth]{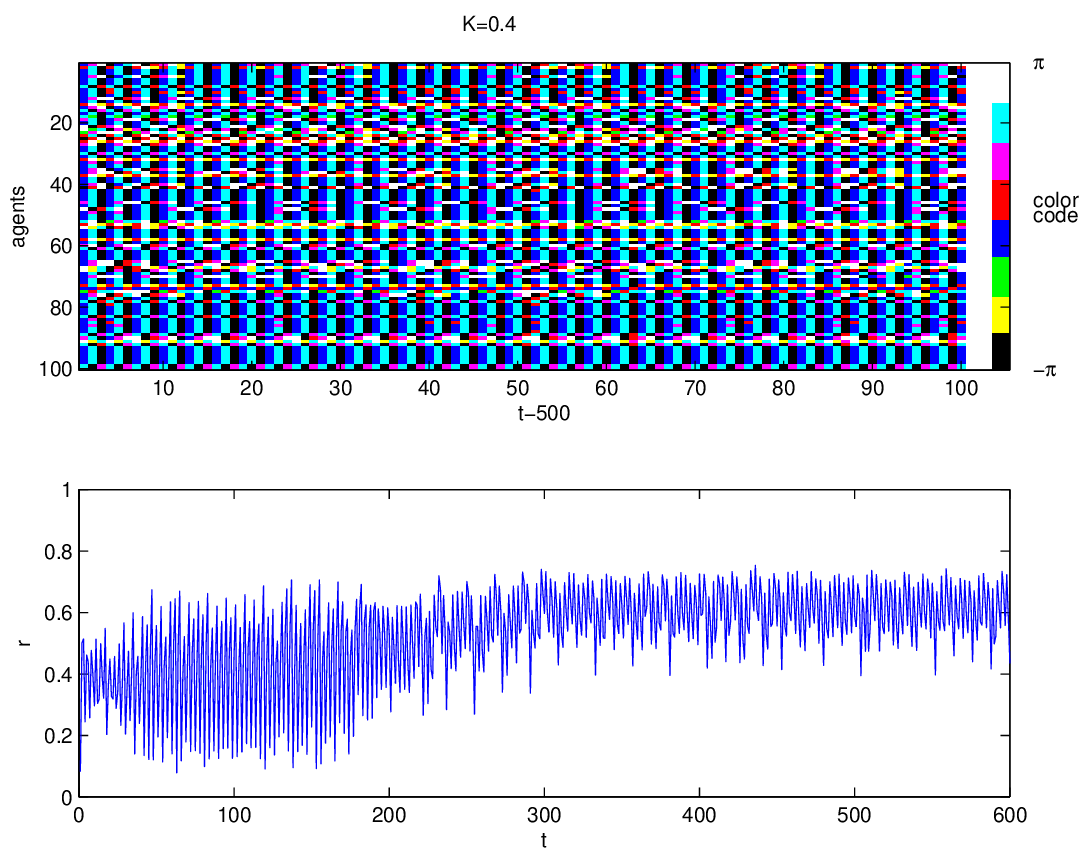}
\caption{Coordinates $x_{i}$ and order parameter $r\left( t\right) $ for $%
K=0.4$}
\label{sync_0_4}
\end{figure}

\begin{figure}[htb]
\centering
\includegraphics[width=0.6\textwidth]{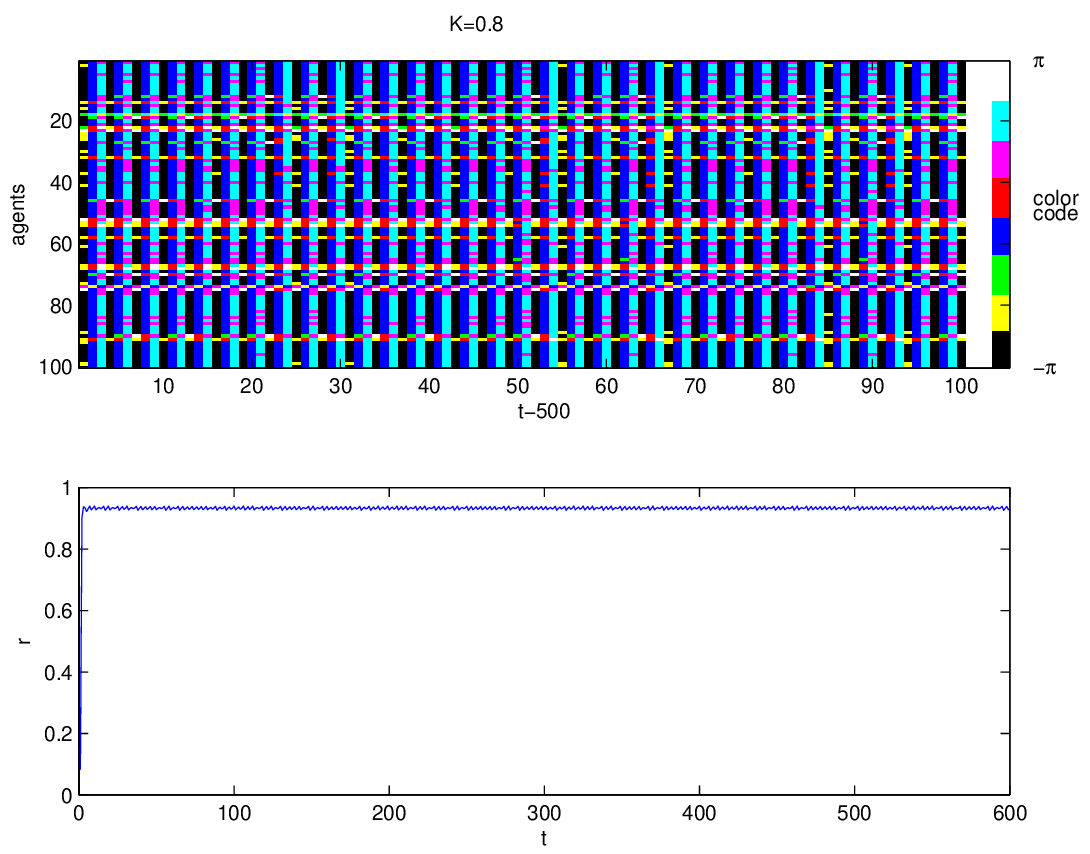}
\caption{Coordinates $x_{i}$ and order parameter $r\left( t\right) $ for $%
K=0.8$}
\label{sync_0_8}
\end{figure}

\begin{figure}[htb]
\centering
\includegraphics[width=0.6\textwidth]{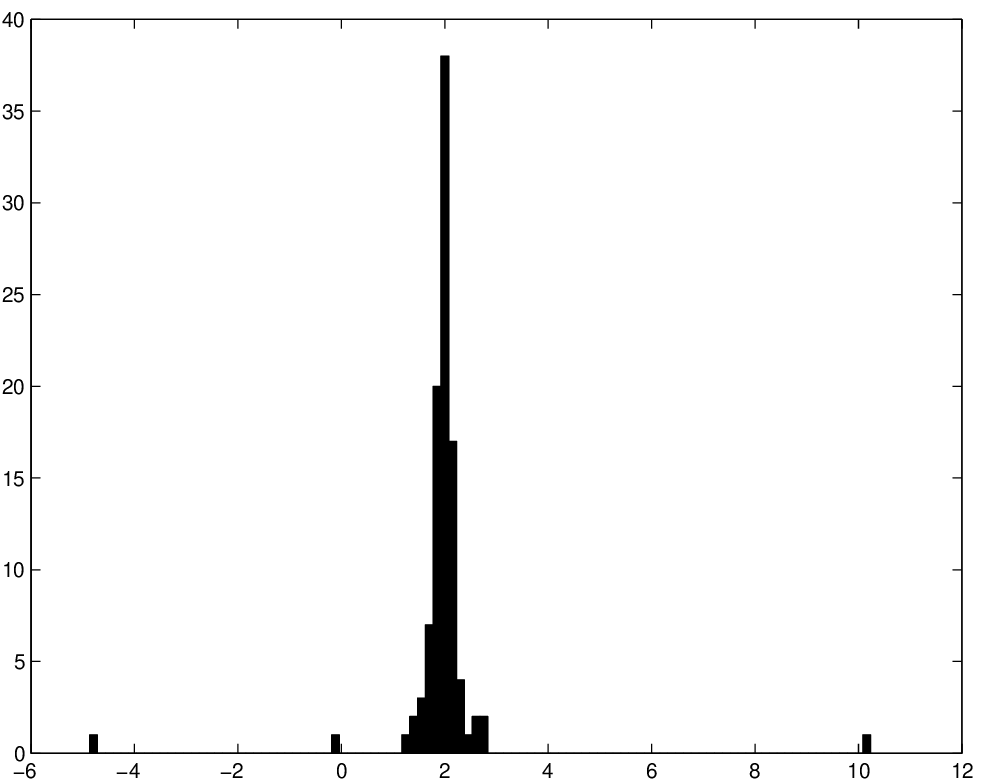}
\caption{A typical distribution of the Cauchy-distributed frequencies $%
\protect\omega _{i}$}
\label{gam}
\end{figure}

In the Figs. \ref{sync_0_2}, \ref{sync_0_4} and \ref{sync_0_8} we display
the results of numerical simulation of the system (\ref{2.1}) with $%
f^{(\infty )}$, $N=100$, $K=0.2$, $K=0.4$ and $K=0.8$. A typical
distribution of the Cauchy-distributed frequencies $\omega _{i}$ is plotted
in Fig.\ref{gam}. We start from random initial conditions and plot the the
color-coded coordinates $x_{i}\left( t\right) $ from $t=500$ to $t=600$. One
sees that for the small $K$ the coordinates seem to be uncorrelated whereas
for larger $K$ a large degree of synchronization is observed. This is also
the information that one obtains from the behavior of the order parameter $%
r\left( t\right) $.

The behavior of the model is similar to Kuramoto's. An important question is
whether synchronization is all there is in the dynamics of interacting
oscillators. In the past, several authors have found, for example,
synchronized cluster formation, before the full synchronization transition.
The simplicity of the present model allows for further exploration of this
question and an useful hint is, as usual, obtained from the computation of
the ergodic parameters \cite{VilelaChaos}. In particular the Lyapunov
spectrum of model (\ref{2.1}), in the $n\rightarrow \infty $ case, may be
obtained exactly.

When $K=0$ there are $N$ neutral directions, that is, the effective
dynamical dimension is $N$ and the Lyapunov spectrum contains $N$ zeros.
However, as soon as $K>0$, the Lyapunov spectrum consists of one isolated
zero and $\log \left( 1-\frac{N}{N-1}K\right) $, ($N-1$)-times. Therefore
although it is only for sufficiently large $K$ that synchronization effects
seem to occur, there are, for any small $K>0$, $N-1$ contracting directions.
The effective dynamical dimension is one for any small $K>0$. As soon as
there is a (positive) interaction between the units, they are, in the
ergodic sense, enslaved to a single collective dynamics. Notice that this is
not a pathology of this model. Numerical simulation of the Kuramoto and
other models also show a drastic reduction of the effective dimension before
the synchronization transition.

The fact that the Lyapunov spectrum of the deformed Kuramoto model in the $%
n\rightarrow \infty $ limit may be obtained exactly, provides important
information on the mechanisms that ocuur before and at synchronization. The
eigenvectors of the Jacobian are%
\begin{equation*}
\left( 
\begin{array}{c}
1 \\ 
1 \\ 
1 \\ 
\vdots \\ 
\vdots \\ 
1%
\end{array}%
\right) ;\left( 
\begin{array}{c}
1 \\ 
-1 \\ 
0 \\ 
\vdots \\ 
\vdots \\ 
0%
\end{array}%
\right) ;\left( 
\begin{array}{c}
1 \\ 
1 \\ 
-2 \\ 
0 \\ 
\vdots \\ 
0%
\end{array}%
\right) ;\cdots ;\left( 
\begin{array}{c}
1 \\ 
1 \\ 
1 \\ 
1 \\ 
\vdots \\ 
-N+1%
\end{array}%
\right)
\end{equation*}%
the first one being associated to the eigenvalue $1$ and all the others to $%
\left( 1-\frac{N}{N-1}K\right) $. Denoting by $x_{i}$ the agents
coordinates, the eigenmodes associated to these eigenvectors are%
\begin{equation*}
Y_{N,p}=\sum_{i=n}^{n+p-1}x_{i}-px_{n+p}
\end{equation*}%
For $K\neq 0$ and before synchronization these modes follow complex periodic
orbits which converge to fixed points when synchronization sets in.
Therefore one finds here two distinct phenomena. The first one is the
dimension reduction at $K=0$ and then the convergence to fixed points of the
eigen-motions associated to the negative Lyapunov exponents. Here the
dimension reduction threshold is quite sharp at $K=0$, but in other models
it is more gradual.

\begin{figure}[htb]
\centering
\includegraphics[width=0.6\textwidth]{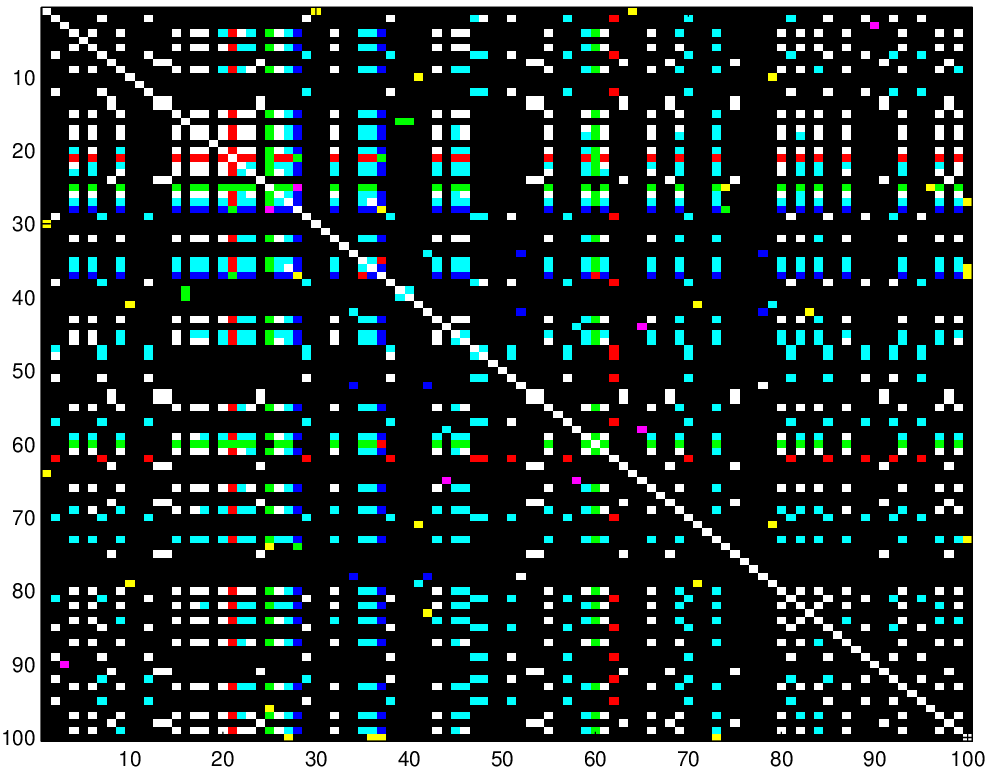}
\caption{The agents' correlation matrix for the deformed Kuramoto model when $n\rightarrow \infty $ and $K=0.2$}
\label{CORR_lin2_K0_2}
\end{figure}

The synchronization order parameter (\ref{2.6}) cannot by itself describe
the strong correlations and dimensional reduction that occur before
synchronization. As shown for the simple model (\ref{2.1}) in the $%
n\rightarrow \infty $ limit, characterization of the correlations may be
obtained by the projections on the eigenvectors of the Lyapunov matrix.
However our aim is to develop general methods that might be applied to any
system when one has no access to its solutions or even to the equations that
generate the time series. Will a correlation measure be sufficient to
unravel all the complexities that arise before synchronization?\ To explore
this possibility we have computed the correlation of the agents dynamics.
That is, we have computed the increments of the coordinates (on the circle),
for a long time interval, and from them the matrix of correlations. A
typical example is shown in Fig.\ref{CORR_lin2_K0_2} with the same colour code as
Fig.\ref{sync_0_2} for positive correlation and black for zero or negative
correlation.

In the figure, at $K=0.2$, one sees many different correlations
of several intensities. However it is not clear, from examination of this
figure, how to characterize the nature of the correlations nor how they
evolve as one approaches the synchronization regime. This has led us to
propose several other methods which were then tested on the models: One is
based on the geometrical characterization of the dynamics, another is
related to dynamical clustering using spectral methods and another one is
based on a version of the notion of conditional Lyapunov exponents.
\clearpage

\subsection{Coupled oscillators with a triangle interaction}

\begin{figure}[htb]
\centering
\includegraphics[width=0.6\textwidth]{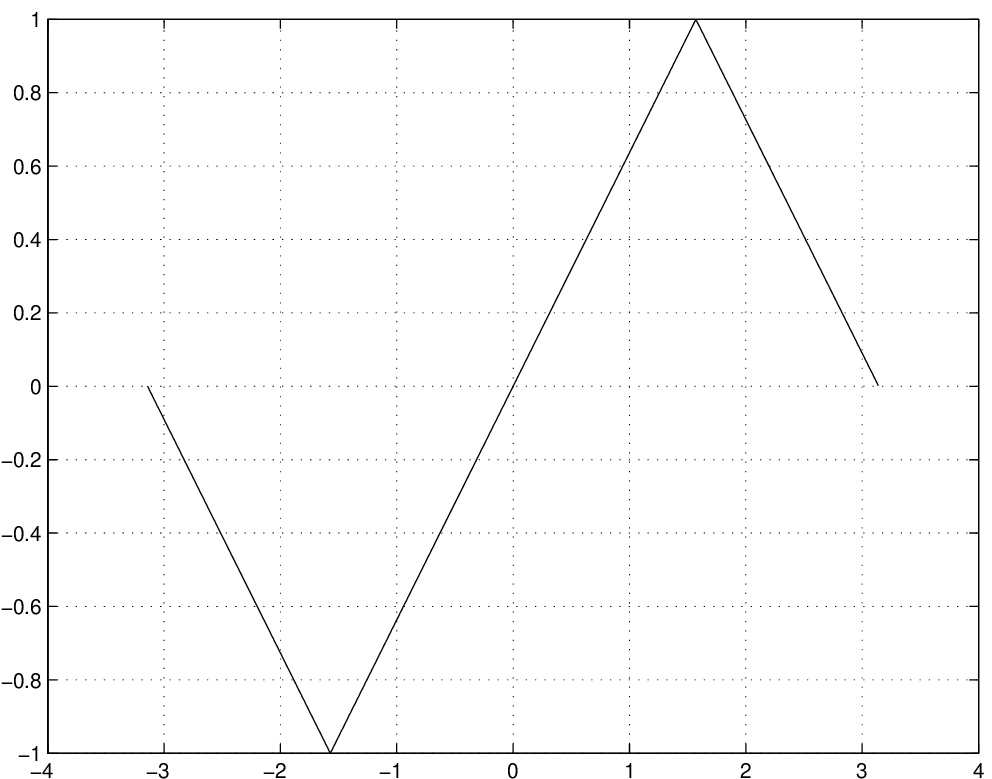}
\caption{The "triangle" function}
\label{triangle}
\end{figure}

Here the dynamical law is%
\begin{equation}
x_{i}\left( t+1\right) =x_{i}\left( t\right) +\omega _{i}+\frac{K}{N-1}%
\sum_{j=1}^{N}g\left( x_{j}-x_{i}\right) \hspace{1cm}(\mathnormal{mod}\pi )
\label{2.7}
\end{equation}%
$g\left( x\right) $ being the function displayed in Fig.\ref{triangle}. The
frequencies $\omega _{i}$ are also assumed to follow a Cauchy distribution.

\begin{figure}[htb]
\centering
\includegraphics[width=0.6\textwidth]{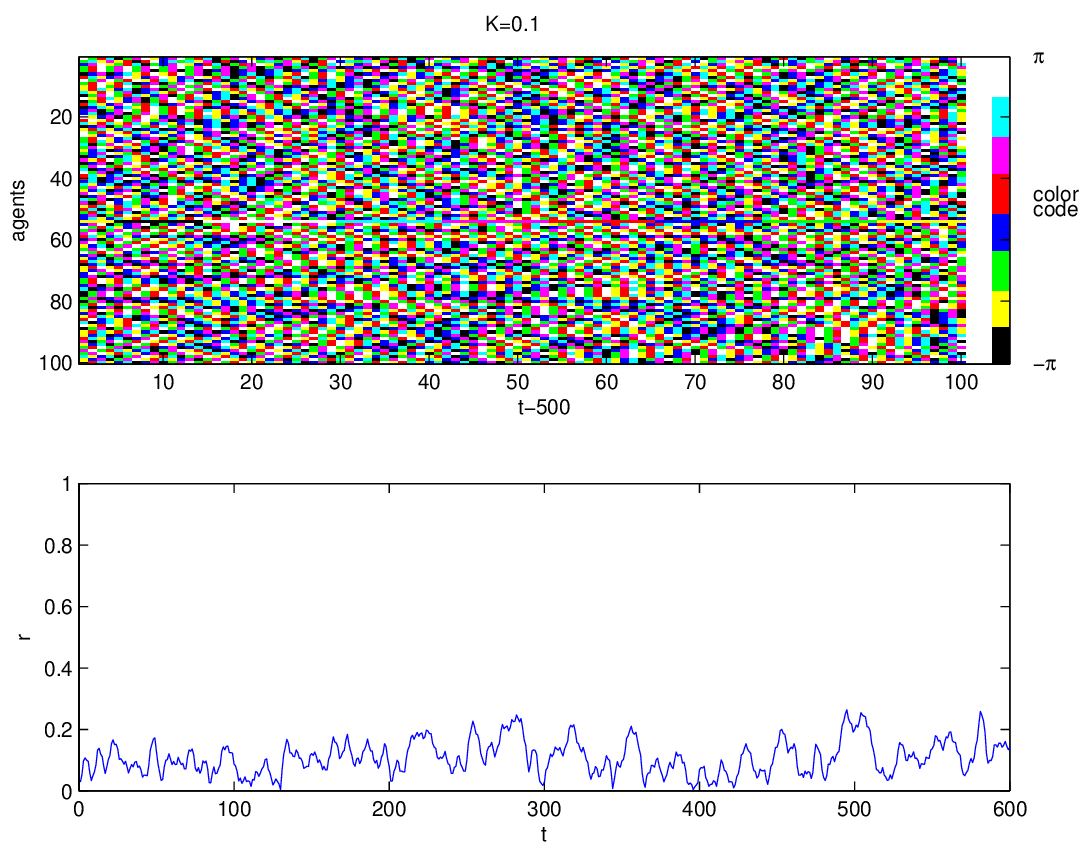}
\caption{Coordinates $x_{i}$ and order parameter $r\left( t\right) $ for $%
K=0.1$ (triangle interaction)}
\label{triang_0_1}
\end{figure}

\begin{figure}[htb]
\centering
\includegraphics[width=0.6\textwidth]{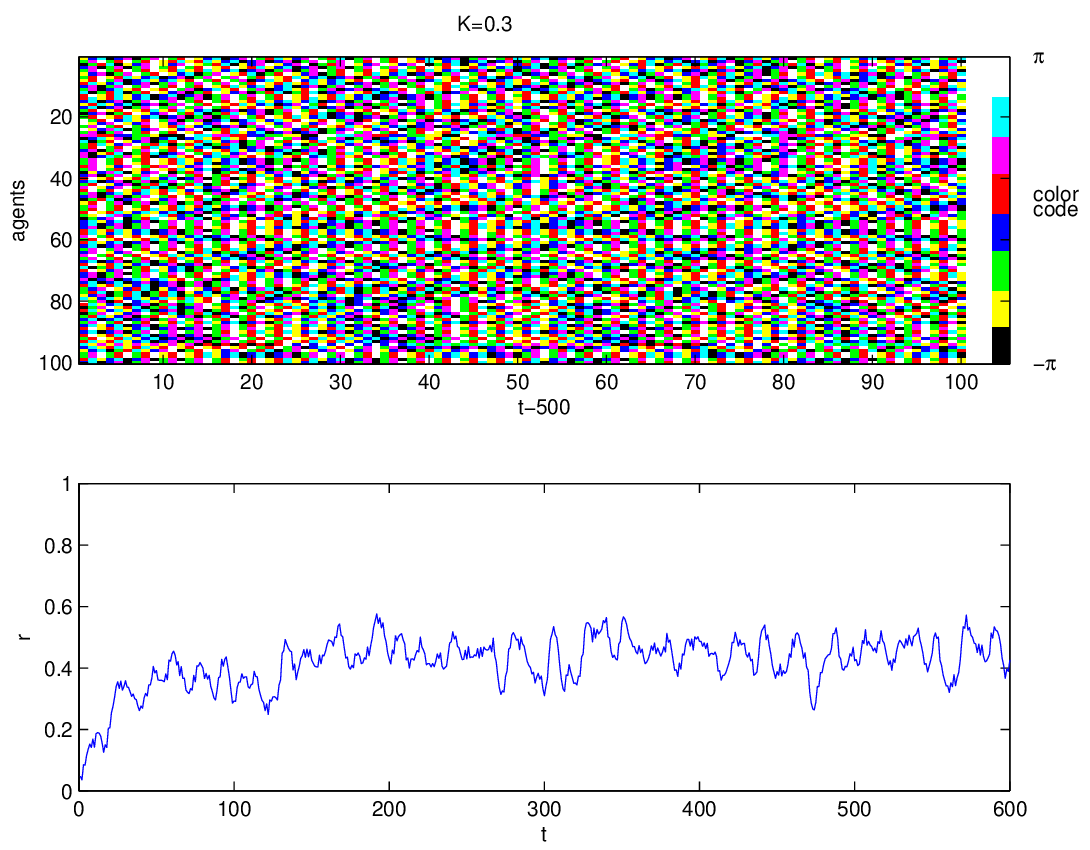}
\caption{Coordinates $x_{i}$ and order parameter $r\left( t\right) $ for $%
K=0.3$ (triangle interaction)}
\label{triang_0_3}
\end{figure}

\begin{figure}[htb]
\centering
\includegraphics[width=0.6\textwidth]{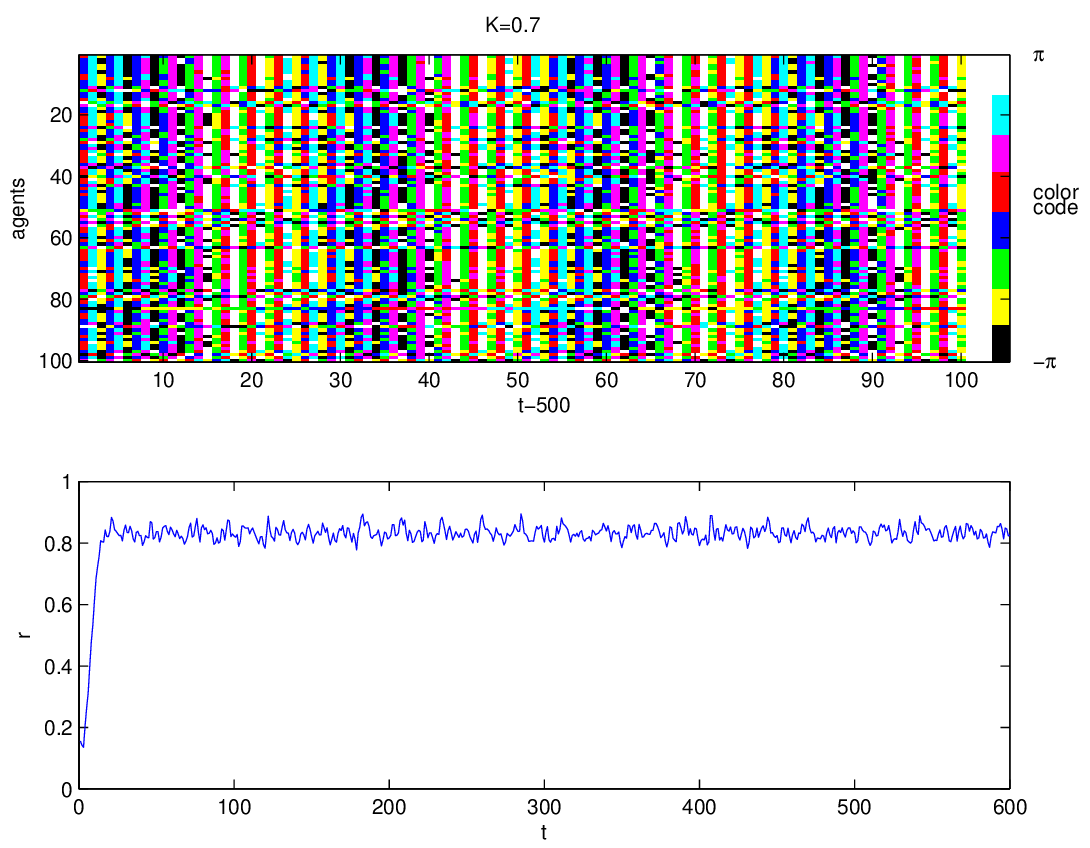}
\caption{Coordinates $x_{i}$ and order parameter $r\left( t\right) $ for $%
K=0.7$ (triangle interaction)}
\label{triang_0_7}
\end{figure}

As in the previous example, for small values of the coupling ($K$) the order
parameter $r$ fluctuates around small values, whereas for large values the
synchronization \ is apparent (Figs.\ref{triang_0_1}, \ref{triang_0_3} and %
\ref{triang_0_7})

However, one sees by computing numerically the Lyapunov spectrum (Fig.\ref%
{Lyaps_triang}), that already for very small $K$ values, instead of $N$
neutral directions there are a number of contracting directions, implying a
reduction in the effective dimension. This is not apparent on the behavior
of the order parameter $r$, emphasizing once more the need to characterize
the correlations that appear before synchronization.

\begin{figure}[htb]
\centering
\includegraphics[width=0.6\textwidth]{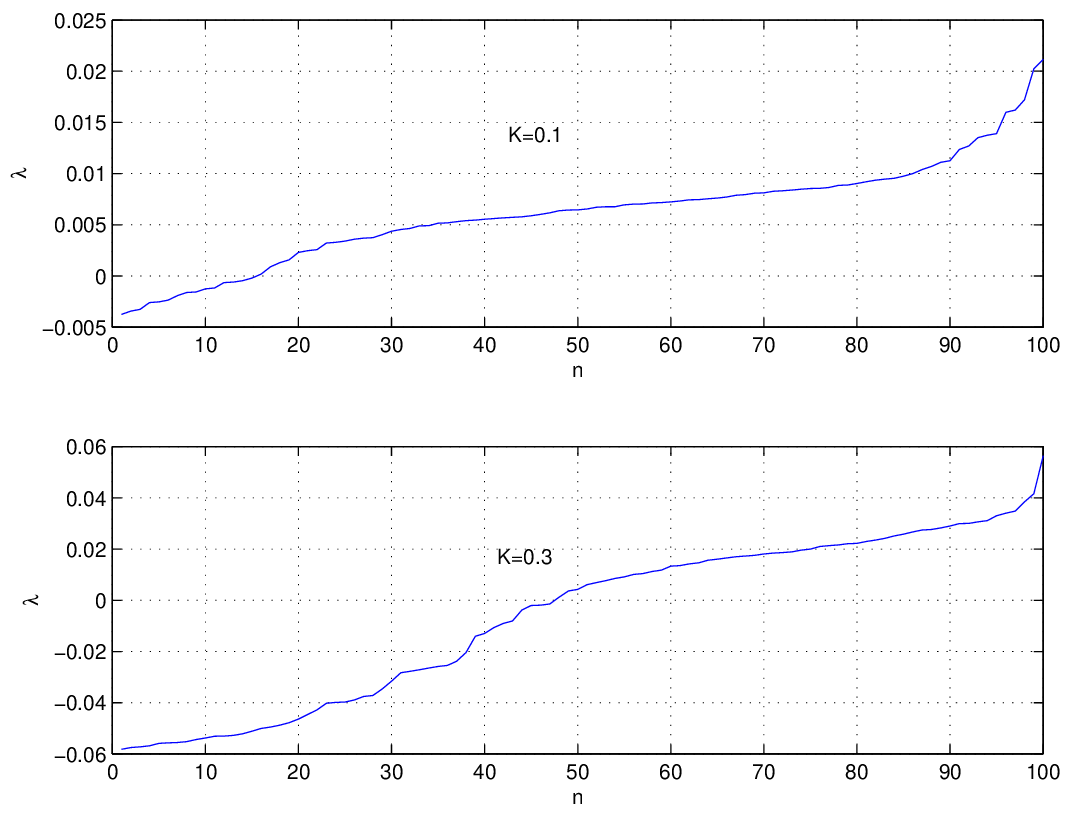}
\caption{Numerically computed Lyapunov spectrum for the triangle interaction
at $K=0.1$ and $K=0.3$}
\label{Lyaps_triang}
\end{figure}
\clearpage

\subsection{A deterministic "integrate and fire" model}

\begin{figure}[htb]
\centering
\includegraphics[width=0.6\textwidth]{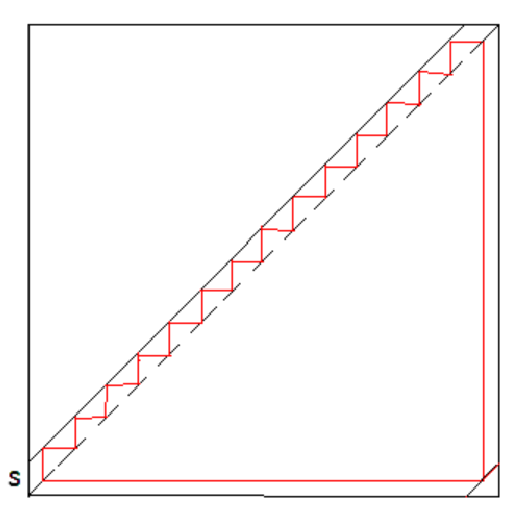}
\caption{The "integrate and fire" free evolution}
\label{inte_fire}
\end{figure}

Our third example is of a different nature from the previous ones. The
dynamics is defined by%
\begin{equation}
x_{i}\left( t+1\right) =x_{i}\left( t\right) +s_{i}+\frac{k}{N-1}\sum_{j\neq
i}\theta \left( x_{j}\left( t-1\right) -x_{j}\left( t\right) -0.4\right) 
\hspace{1cm}\left( \mathnormal{mod}1\right)  \label{2.8}
\end{equation}%
$\theta $ being the function%
\begin{equation*}
\left\{ 
\begin{array}{ccc}
x>0 &  & \theta \left( x\right) =1 \\ 
x\leq 0 &  & \theta \left( x\right) =0%
\end{array}%
\right.
\end{equation*}%
The free evolution of each unit is a slow increase during many time steps
followed by a jump (Fig.\ref{inte_fire}).

\begin{figure}[htb]
\centering
\includegraphics[width=0.6\textwidth]{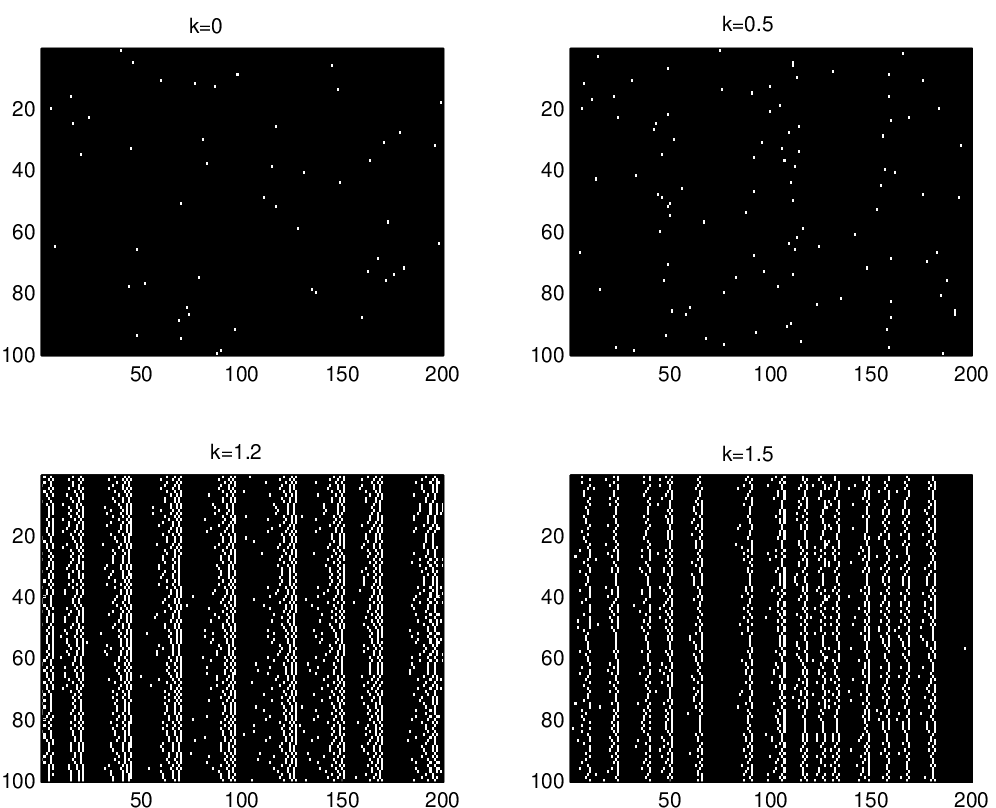}
\caption{Spiking patterns for different coupling values. 200 time steps, 100
units.}
\label{firings_inte_fire}
\end{figure}

This jump is, by a neuron analogy \cite{Aguirre}, interpreted as a spike and
the interaction with the other units occurs only when they spike. In the Fig.%
\ref{firings_inte_fire} we display the time evolution of the spiking units
obtained for $k=0,0.5,1.2$ and $1.5$. The simulations are run from random
initial conditions in the unit interval and the $s_{i}%
{\acute{}}%
s$ are also chosen at random.

As the coupling increases one sees an increase in the spiking rate but not
special coordination between the firing times. However, above around $k=1$ a
distinct clustering of the spiking patterns is clearly observed. How these
correlations may be characterized will later be seen.

\clearpage
\section{Tools to characterize correlations in collective dynamics}

In this section we describe some tools which may, qualitative and
quantitatively characterize correlations in collective dynamics. The
emphasis is, as stated before, on the characterization of the nature of the
correlations that occur before synchronization sets in or even in systems
that never synchronize.

\subsection{The geometry of the dynamics}

Given a set of $N$ time series one defines a distance between each pair. One
possibility is to consider the Euclidean distance%
\begin{equation}
d_{ij}=\frac{1}{T-t_{0}}\sqrt{\sum_{t=t_{0}}^{T}\left( x_{i}\left( t\right)
-x_{j}\left( t\right) \right) ^{2}}  \label{3.1}
\end{equation}%
Then, using the technique of multidimensional scaling (MDS), imbed the $N$
time series as points in an Euclidean space. MDS begins with a $N\times N$
distance matrix $D=\left\{ d_{ij}\right\} $ and the aim is to find a
configuration of points in a $p-$dimensional space such that the coordinates
of the points yield a Euclidean distance matrix with elements which are as
close as possible to the distances in the original distance matrix (but not
exactly the same if the original distances are not Euclidean).

We proceed as follows: denote by $Y$ the matrix of coordinates in the
embedding $p-$dimensional Euclidean space%
\begin{equation}
Y=\left( 
\begin{array}{ccccc}
y_{11} & y_{12} & \cdots & \cdots & y_{1p} \\ 
y_{21} & y_{22} & \cdots & \cdots & y_{2p} \\ 
\vdots & \vdots & \vdots & \vdots & \vdots \\ 
y_{N1} & y_{N2} & \cdots & \cdots & y_{Np}%
\end{array}%
\right)  \label{3.2}
\end{equation}%
and consider the following decomposition of the squared distance matrix%
\begin{equation}
d_{ij}^{2}=\left\vert \overset{\longrightarrow }{y_{i}}-\overset{%
\longrightarrow }{y_{j}}\right\vert ^{2}=b_{ii}+b_{jj}-2b_{ij}  \label{3.4}
\end{equation}%
Then, the elements of the $N\times N$ matrix $B$%
\begin{equation}
B=YY^{T}  \label{3.3}
\end{equation}%
are recovered from%
\begin{equation}
b_{ij}=-\frac{1}{2}\left\{ d_{ij}^{2}-\frac{1}{n}\left(
\sum_{j=1}^{n}d_{ij}^{2}+\sum_{i=1}^{n}d_{ij}^{2}-\frac{1}{n}%
\sum_{i,j=1}^{n}d_{ij}^{2}\right) \right\}  \label{3.5}
\end{equation}%
where by a translation of the origin in $\mathbb{R}^{p}$ one makes$\
\sum_{i=1}^{N}y_{ik}=0$ for all $k$.

One diagonalizes the matrix $B$ reconstructed by (\ref{3.5})%
\begin{equation}
B=V\Lambda V^{T}  \label{3.6}
\end{equation}%
with $\Lambda =\left( \lambda _{1}\cdots \lambda _{n}\right) $ ($\lambda
_{1}\geq \lambda _{2}\geq \cdots \geq \lambda _{N}$) being the diagonal
matrix of eigenvalues and $V=\left[ V_{1},\cdots ,V_{N}\right] $ the matrix
of normalized eigenvectors. Whenever the dimension $p$ of the imbedding
space is smaller than $N$ the rank of $B$ is $p$ (with the last $N-p$
eigenvalues being zero). One may write%
\begin{equation}
B=V^{\ast }\Lambda ^{\ast }V^{\ast T}  \label{3.7}
\end{equation}%
where $V^{\ast }$ contains the first $p$ eigenvectors and $\Lambda ^{\ast }$
the first $p$ eigenvalues. Then a solution for $Y$ is $Y=V^{\ast }\Lambda
^{\ast 1/2}$.

When the input distance matrix is not Euclidean, the matrix $B$ is not
positive-definite. In such case, some of the eigenvalues of $B$ will be
negative and correspondingly some coordinate values will be complex numbers.
If $B$ has only a small number of small negative eigenvalues, it is still
possible to use the eigenvectors associated with the $p$ largest positive
eigenvalues.

For the time series case, after the Euclidean embedding of the orbits is
done, one obtains a cloud of points (a point for each orbit). The shape and
effective dimension of the cloud is obtained by reducing the coordinates to
the center of mass and computing the inertial tensor%
\begin{equation}
T_{ij}=\sum_{k=1}^{N}y_{i}(k)y_{j}(k)  \label{3.9}
\end{equation}%
Let $\lambda \left( T\right) $ be the eigenvalues of $T$. Once the
eigenvalues $\left\{ \lambda _{k}\right\} $ and eigenvectors $\left\{
V_{k}\right\} $ of $T$ are found, the relevant quantities, to characterize
the correlations, are the projections $\left( x_{i},V_{k}\right) $\ of the
coordinate vectors on the eigenvectors, in particular on those associated to
the largest eigenvalues.

\subsection{Dynamical clustering}

Here one wants to develop a tool to detect the dynamical communities that
emerge from the interaction. For this purpose the relevant quantities
characterizing the dynamics of each agent are the coordinate increments%
\begin{equation}
\Delta _{i}\left( t\right) =x_{i}\left( t\right) -x_{i}\left( t-1\right)
\label{4.1}
\end{equation}%
which may be used to find a dynamical distance of the agents%
\begin{equation}
d_{ij}=\sqrt{\sum_{t=1}^{T}\left\vert \Delta _{i}\left( t\right) -\Delta
_{j}\left( t\right) \right\vert ^{2}}  \label{4.2}
\end{equation}

From the distances one defines an adjacency matrix%
\begin{equation}
A_{ij}=\exp \left( -\beta (d_{ij}-d_{\min })\right)  \label{4.3}
\end{equation}%
a degree matrix%
\begin{equation}
\left( G\right) _{ii}=\sum_{j\neq i}A_{ij}  \label{4.4}
\end{equation}%
and a Laplacian matrix%
\begin{equation}
L=G-A  \label{4.5}
\end{equation}%
The lowest eigenvalues in the $L-$spectrum provide information on the
dynamical communities insofar as they minimize the RatioCut of $K$
communities \cite{vonLuxburg}%
\begin{equation}
\text{RatioCut}\left( C_{1},...,C_{K}\right) =\frac{1}{2}\sum_{k=1}^{K}\frac{%
W\left( C_{k},\overline{C_{k}}\right) }{\left\vert C_{k}\right\vert }
\label{4.6}
\end{equation}%
with $W\left( C_{k},\overline{C_{k}}\right) =\sum_{i\in C_{k},j\in \overline{%
C_{k}}}A_{ij}$ being the sum of the external connections of the community $%
C_{k}$, and $\left\vert C_{k}\right\vert $ the number of elements in the $%
C_{k}$ community.

\subsection{The conditional Lyapunov spectrum}

An issue of some relevance in multi-agent systems is to compare the view
that each agent has of its dependence on the dynamics of the other agents
with the actual dependence on the dynamics of the whole network. This is
captured by the notion of \textit{conditional exponents}. Conditional
exponents, first introduced by Pecora and Carroll \cite{Pecora1} in their
study of synchronization of chaotic systems, have been shown to be good
ergodic invariants \cite{VilelaCondExp}, playing an important role as
self-organization parameters \cite{VilelaSelfOrg}. The conditional exponents
are computed in a way similar to the Lyapunov exponents but with each agent
taking into account only its neighbors, not the whole system. However for
the time average required for the calculation of the ergodic invariants, the
actual global dynamics is used.

For a system with the neighborhood degree characterized by the adjacency
matrix, the calculation of the conditional exponents spectrum is equivalent
to the computation of the Jacobian of a modified dynamics where the
interaction is weighed by the proximity of the agents (that is, by the
adjacency matrix). Nevertheless the Jacobian is averaged over the orbits of
the actual dynamics. For example, for the interacting oscillators of the
deformed Kuramoto model, the Jacobian would be computed for a fictitious
dynamics%
\begin{equation}
x_{i}\left( t+1\right) =x_{i}\left( t\right) +\omega _{i}+\frac{K}{N-1}%
\sum_{j=1}^{N}A_{ij}\pi f^{(n)}\left( x_{j}-x_{i}\right)  \label{5.1}
\end{equation}%
The integrated difference of the the conditional and the Lyapunov spectrum
is an important parameter to characterize the correlated dynamics.

Another promising technique to characterize the correlations occurring
before synchronization has been developed by Lopez and Rodriguez \cite%
{Rodriguez} who, by considering the Hilbert transform of the coupled time
series, obtain an evolving phase and then compute the entropy of the phase
distribution. We will not deal here with this technique and refer to \cite%
{Rodriguez} for details.

How the above techniques do indeed provide information on the correlations
of the collective dynamics will be clear by their application to the models
described in Section 2.

\section{Illustration of the tools on the model interactions}

\subsection{The deformed Kuramoto model}

\subsubsection{The geometry of the dynamics}

We have applied the geometrical technique to the model (\ref{2.1}), the
distance of agent $i$ to agent $j$ being the sum of the distances on the
circle on the last 100 time steps. Embedding each $100-$times orbit as
points in Euclidean space and using MDS, the eigenvalues $\lambda \left(
B\right) $ of the $B$ matrix were obtained.

The coordinates of the imbedded dynamics are then reduced to the center of
mass and the inertial tensor is computed,%
\begin{equation}
T_{ij}=\sum_{k=1}^{N}y_{i}(k)y_{j}(k)  \label{6.1}
\end{equation}%
the eigenvalues $\left\{ \lambda _{k}\left( T\right) \right\} $ being the
eigenvalues of $T$ and $\left\{ V_{k}\right\} $ its eigenvectors.

\begin{figure}[htb]
\centering
\includegraphics[width=0.6\textwidth]{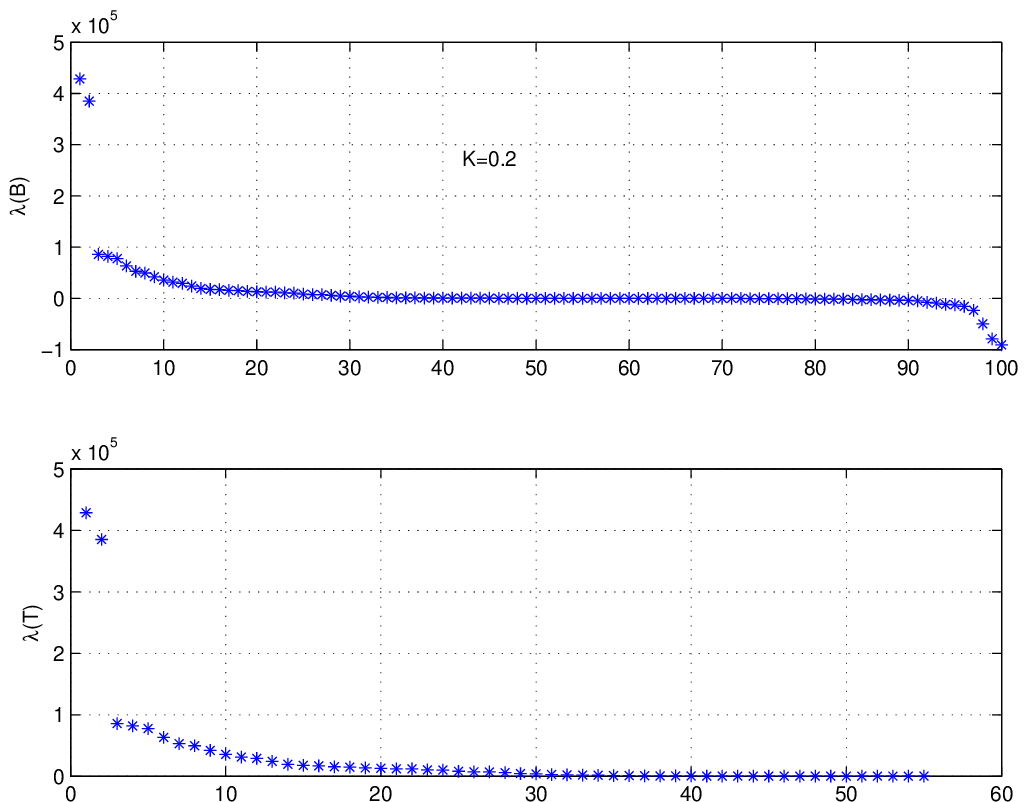}
\caption{Eigenvalues of the $B$ (Eq.\protect\ref{3.3}) and $T$ (Eq.\protect
\ref{3.9}) matrices for $K=0.2$}
\label{LambdaBT_0_2}
\end{figure}

\begin{figure}[htb]
\centering
\includegraphics[width=0.6\textwidth]{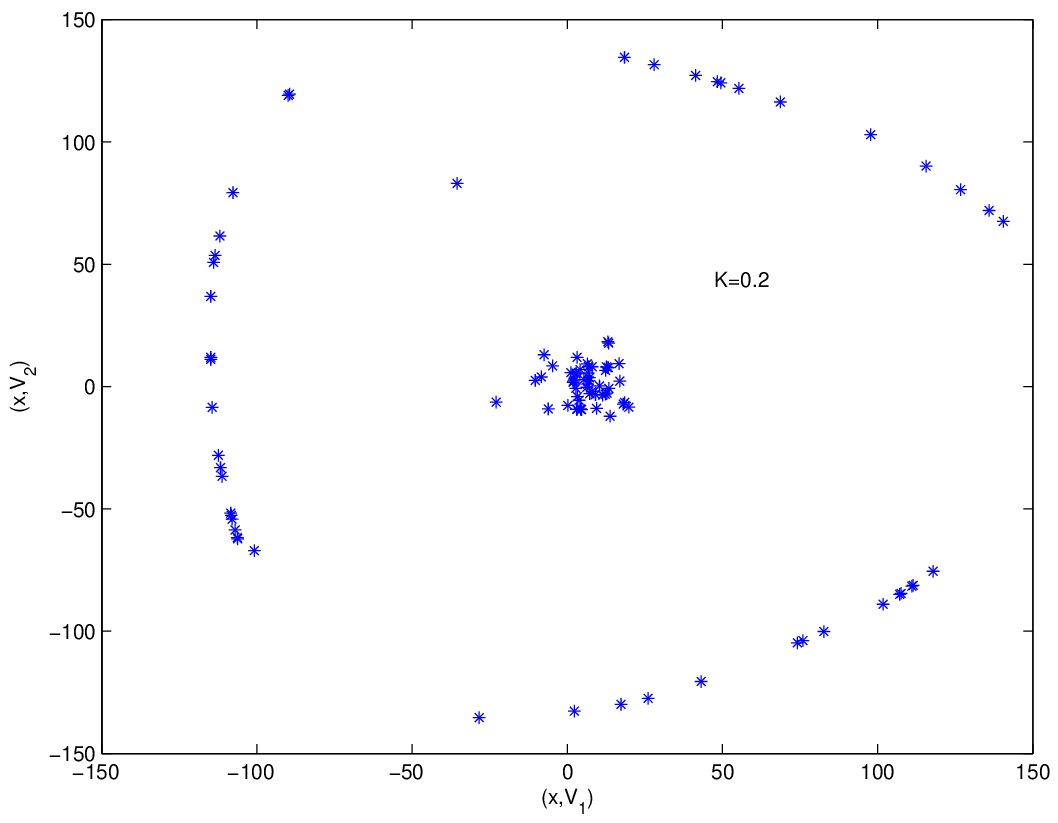}
\caption{Projection of the dynamics on the first and second eigenvectors for 
$K=0.2$}
\label{geomproj_0_2}
\end{figure}

\begin{figure}[htb]
\centering
\includegraphics[width=0.6\textwidth]{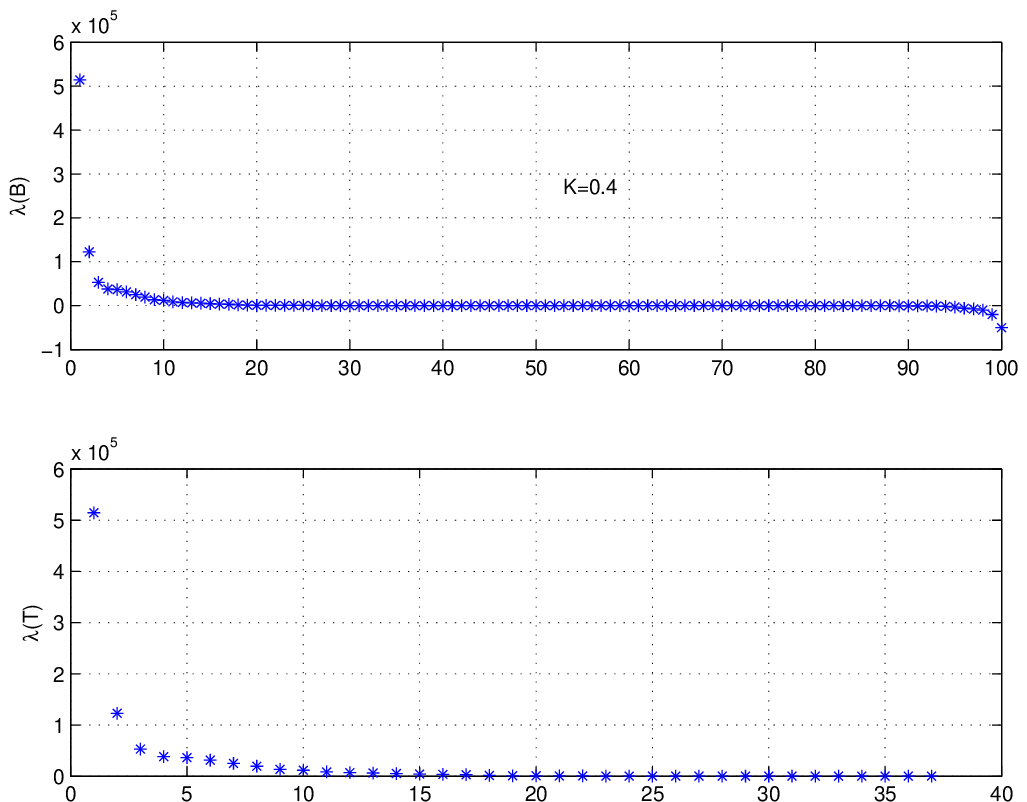}
\caption{Eigenvalues of the $B$ (Eq.\protect\ref{3.3}) and $T$ (Eq.\protect
\ref{3.9}) matrices for $K=0.4$}
\label{LambdaBT_0_4}
\end{figure}

\begin{figure}[htb]
\centering
\includegraphics[width=0.6\textwidth]{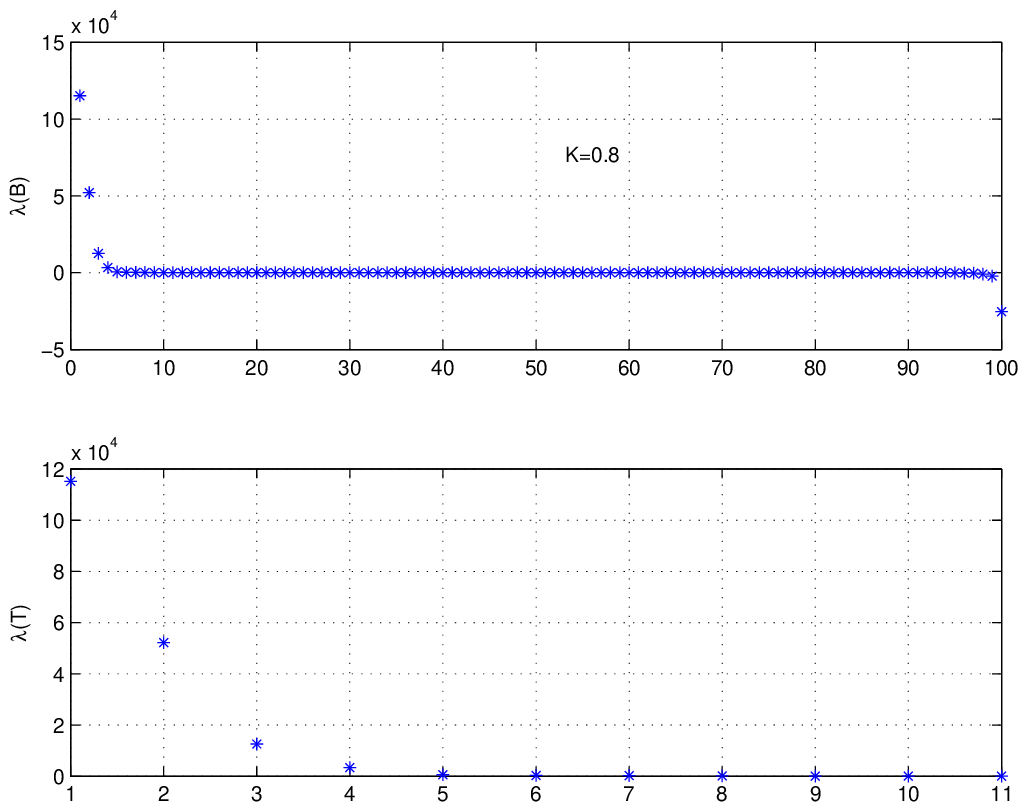}
\caption{Eigenvalues of the $B$ (Eq.\protect\ref{3.3}) and $T$ (Eq.\protect
\ref{3.9}) matrices for $K=0.8$}
\label{LambdaBT_0_8}
\end{figure}

\begin{figure}[htb]
\centering
\includegraphics[width=0.6\textwidth]{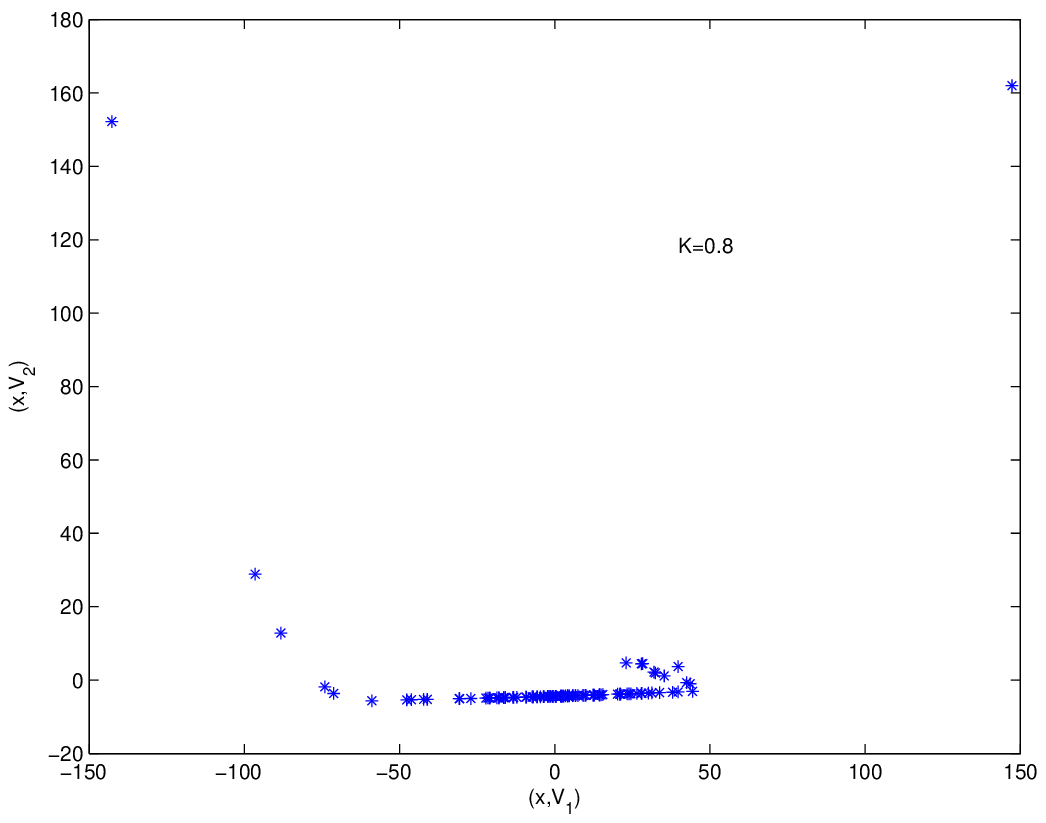}
\caption{Projection of the dynamics on the first and second eigenvectors for 
$K=0.8$}
\label{geomproj_0_8}
\end{figure}

Figs. \ref{LambdaBT_0_2} and \ref{geomproj_0_8} show the results of the
geometrical analysis for the dynamics of the model (\ref{2.1}). Figs. \ref%
{LambdaBT_0_2}, \ref{LambdaBT_0_4} and \ref{LambdaBT_0_8} show the
eigenvalues of the $B$ and $T$ matrices and Figs. \ref{geomproj_0_2}, \ref%
{geomproj_0_4} and \ref{geomproj_0_8} the projection of the dynamics on the
first and second eigenvectors of $T$. Of particular interest is the fast
reduction in the geometrical dimension of the dynamics, as measured by the
fast convergence to zero of the $\lambda \left( T\right) $ eigenvalues, for $%
K\neq 0$. The whole dynamics seems to be approximately embedded in a
two-dimensional subspace. Therefore, the projections on the first two
(dominant) eigenvectors which display very distinct organized patterns,
exhibit the strong correlations that already exist before synchronization
sets in.

\begin{figure}[htb]
\centering
\includegraphics[width=0.6\textwidth]{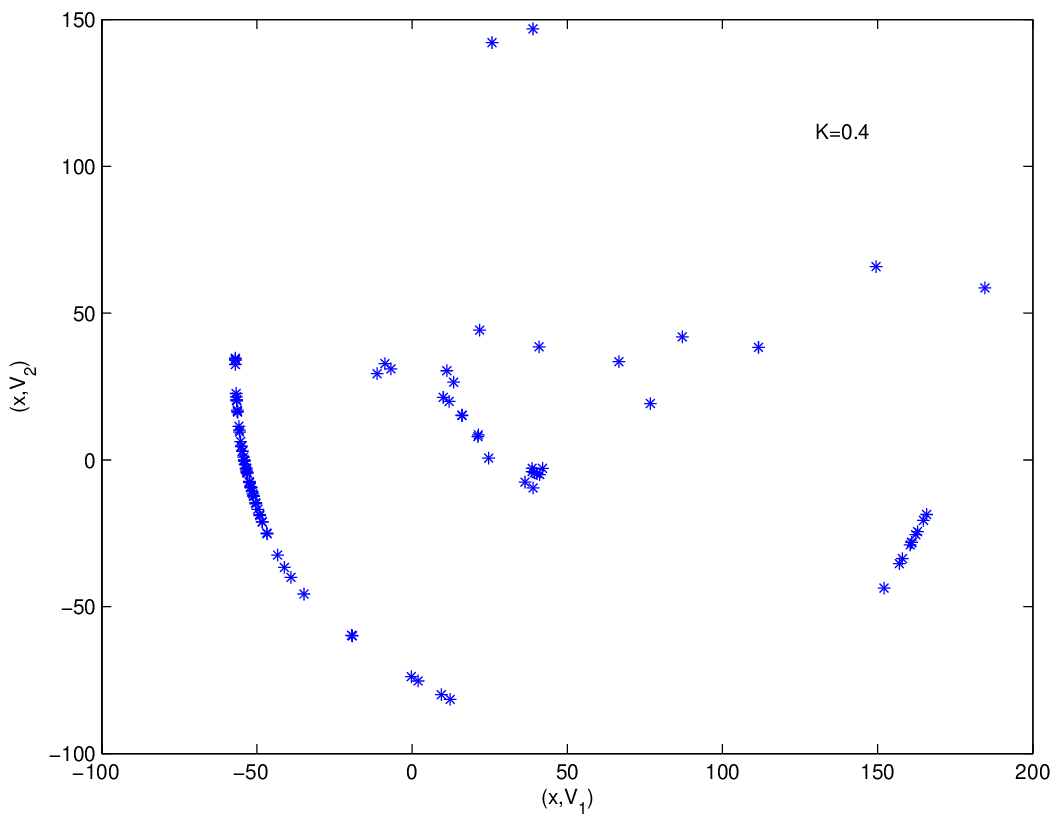}
\caption{Projection of the dynamics on the first and second eigenvectors for 
$K=0.4$}
\label{geomproj_0_4}
\end{figure}

The projection of the embedded coordinates $\left\{ x_{i}\right\} $ on the
eigenvectors $V_{k}$ associated to the largest eigenvalues of $T$ may be
considered as the new order parameters that characterize the correlations
that occur before synchronization. Of interest are also the parameters $%
P_{k}=\sum_{i=1}^{N}\left\vert \left( x_{i},V_{k}\right) \right\vert $.

\subsubsection{Dynamical clustering}

\begin{figure}[htb]
\centering
\includegraphics[width=0.6\textwidth]{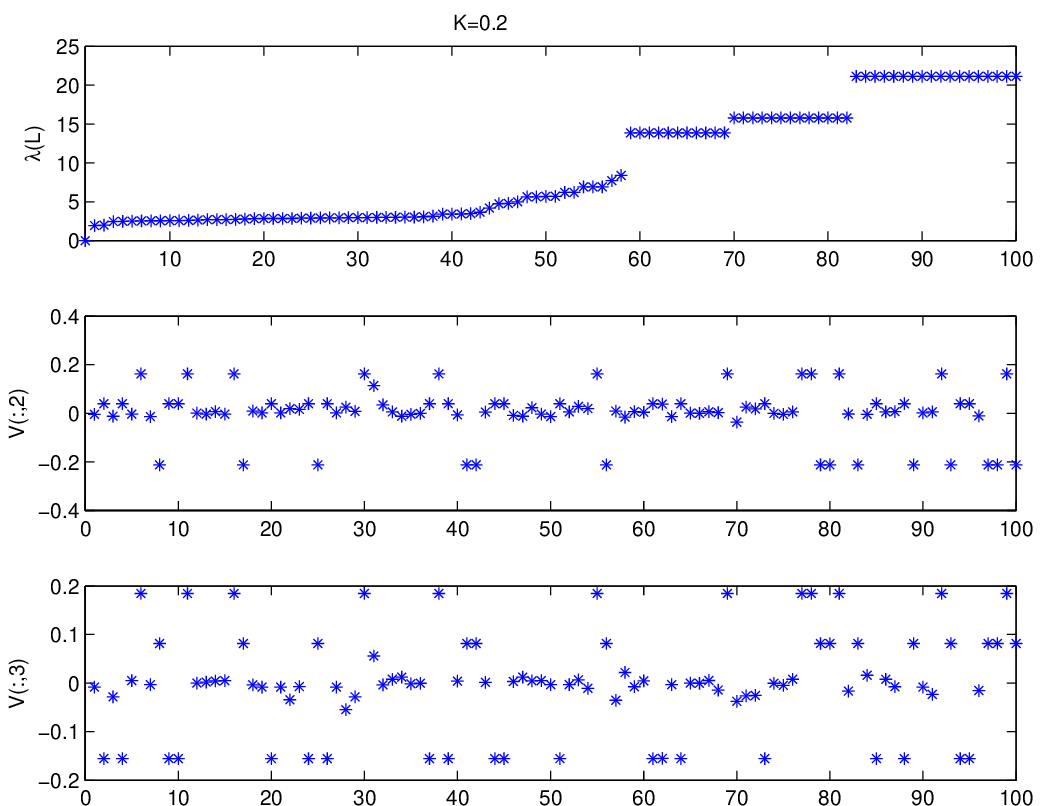}
\caption{The spectrum of the Laplacian matrix $L$ and the second and third
eigenvectors ($K=0.2$)}
\label{lspectr_0_2}
\end{figure}

\begin{figure}[htb]
\centering
\includegraphics[width=0.6\textwidth]{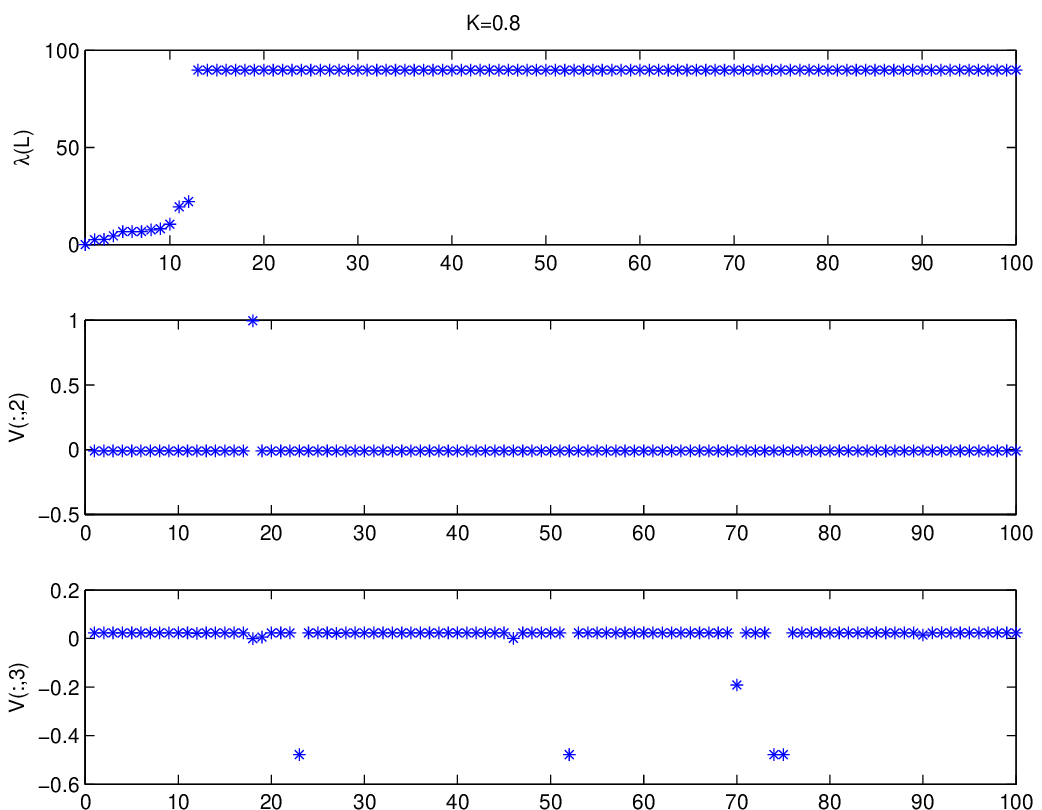}
\caption{The spectrum of the Laplacian matrix $L$ and the second and third
eigenvectors ($K=0.8$)}
\label{lspectr_0_8}
\end{figure}

Distances and adjacency matrices were computed from the coordinate
increments (Eqs.\ref{4.1} to \ref{4.4}). In Figs. \ref{lspectr_0_2}, \ref%
{lspectr_0_4} and \ref{lspectr_0_8} we have plotted the spectrum of the
Laplacian matrix $L$ as well as the structure of the second and third
eigenvectors to show the nature of the dominant communities.

\begin{figure}[htb]
\centering
\includegraphics[width=0.6\textwidth]{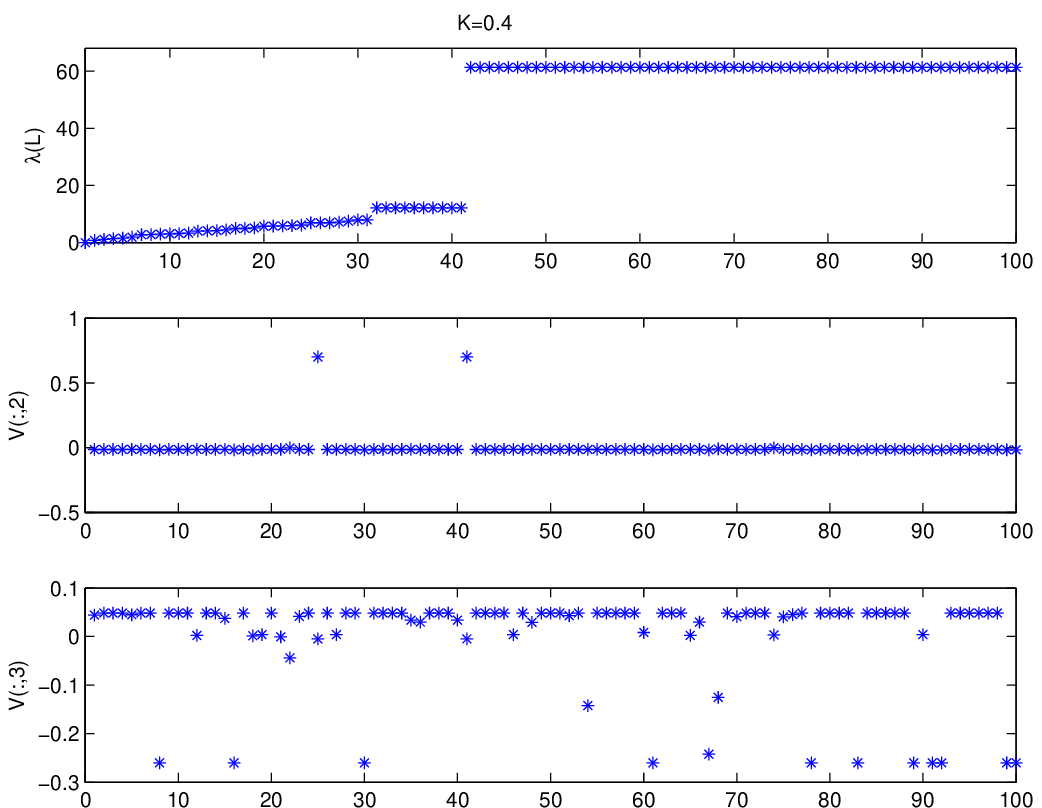}
\caption{The spectrum of the Laplacian matrix $L$ and the second and third
eigenvectors ($K=0.4$)}
\label{lspectr_0_4}
\end{figure}

\subsubsection{The conditional exponents spectrum}

As explained before, the conditional exponents are obtained from the
Jacobian weighed by the agents proximity (that is, by the adjacency matrix)
averaged over the orbits of the actual dynamics. For the deformed Kuramoto
model, the Jacobian is computed for a fictitious dynamics%
\begin{equation}
x_{i}\left( t+1\right) =x_{i}\left( t\right) +\omega _{i}+\frac{K}{N-1}%
\sum_{j=1}^{N}A_{ij}\pi f^{(n)}\left( x_{j}-x_{i}\right)  \label{7.1}
\end{equation}%
The adjacency matrix that is used is the same that was derived in the
previous subsection (4.1.2).

\begin{figure}[htb]
\centering
\includegraphics[width=0.6\textwidth]{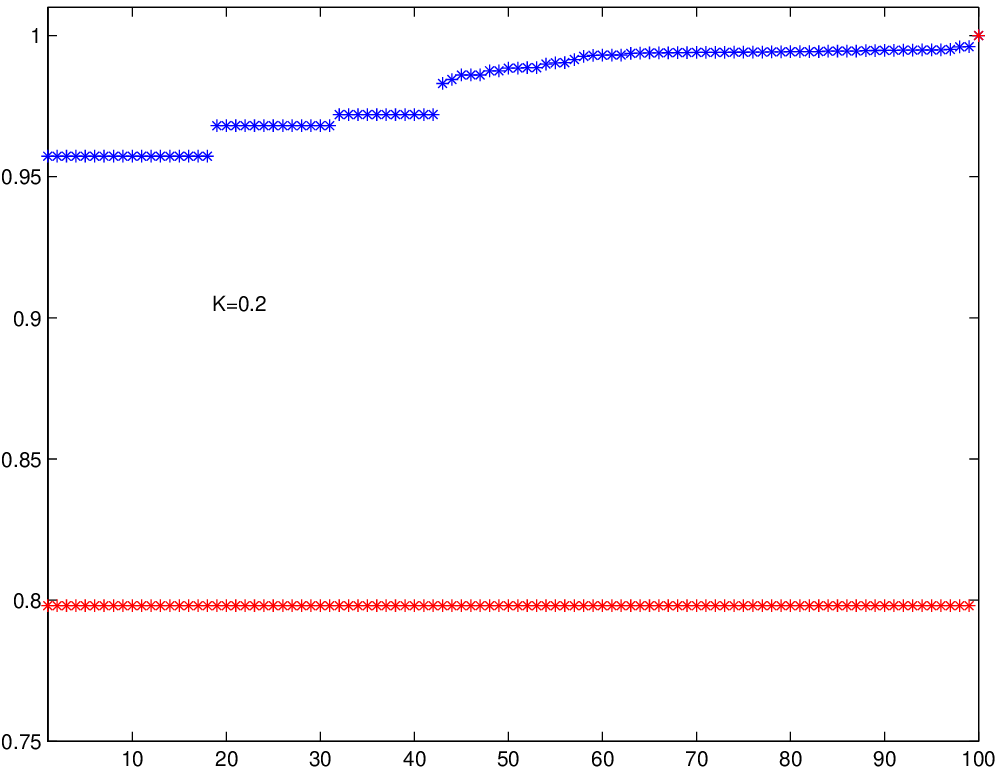}
\caption{Conditional (blue) versus Lyapunov (red) numbers ($K=0.2$)}
\label{jspectr_0_2}
\end{figure}

\begin{figure}[htb]
\centering
\includegraphics[width=0.6\textwidth]{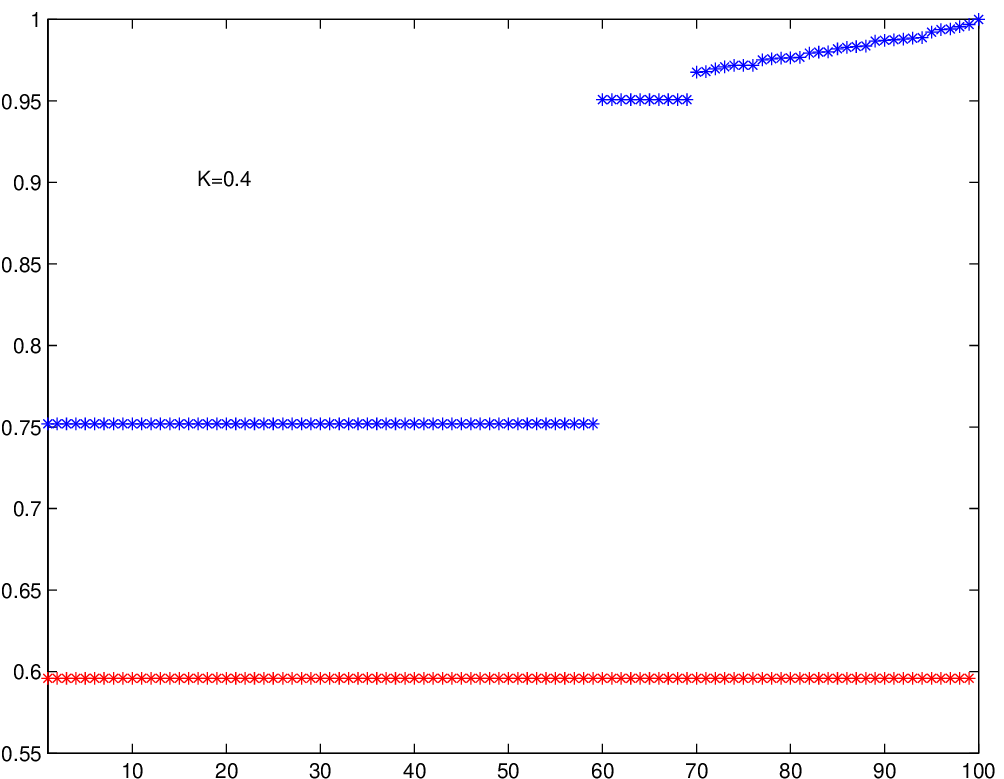}
\caption{Conditional (blue) versus Lyapunov (red) numbers ($K=0.4$)}
\label{jspectr_0_4}
\end{figure}

\begin{figure}[htb]
\centering
\includegraphics[width=0.6\textwidth]{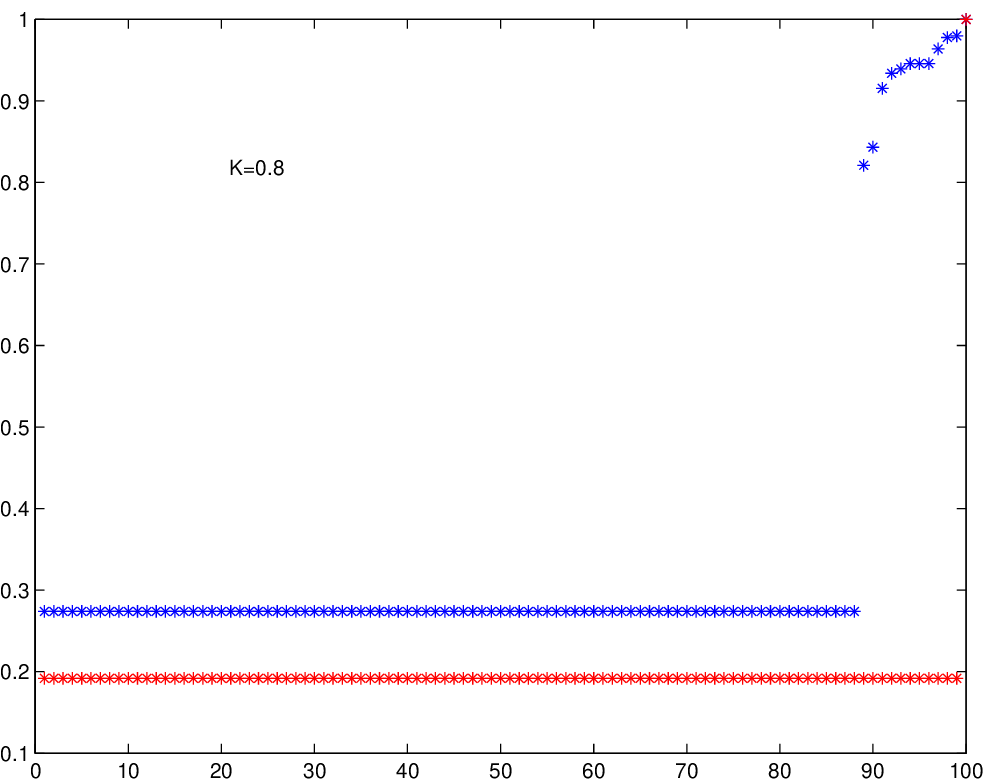}
\caption{Conditional (blue) versus Lyapunov (red) numbers ($K=0.8$)}
\label{jspectr_0_8}
\end{figure}

In Figs. \ref{jspectr_0_2}, \ref{jspectr_0_4} and \ref{jspectr_0_8} the the
spectrum of the Lyapunov number $\left( \mu _{i}=e^{\lambda _{i}}\right) $
of the system (\ref{2.1}) is compared with the conditional number $\left(
\mu _{i}^{C}=e^{\lambda _{i}^{C}}\right) $ spectrum for $K=0.2$, $0.4$ and $%
0.8$.

One sees that for small coupling the conditional number spectrum is still
close to the spectrum of the uncoupled system, meaning that the "perception"
of the agents is very close to a situation where their dynamics looks as a
free dynamics, although in fact it is already fully correlated, as evidenced
by the Lyapunov spectrum. As the coupling increases the conditional number
spectrum becomes closer and closer to the Lyapunov spectrum. The integrated
difference of the two spectra is an important parameter to characterize the
correlated dynamics.
\clearpage

\subsection{The triangle interaction model}

For the triangle interaction, the dynamical dimension reduction is not as
dramatic as in the deformed Kuramoto model, as is evident from the behavior
of its Lyapunov spectrum (Fig. \ref{Lyaps_triang}). Therefore one expects
the correlations to develop at a slower pace as the coupling ($K$) increases.

\subsubsection{The geometry of the dynamics}

\begin{figure}[htb]
\centering
\includegraphics[width=0.6\textwidth]{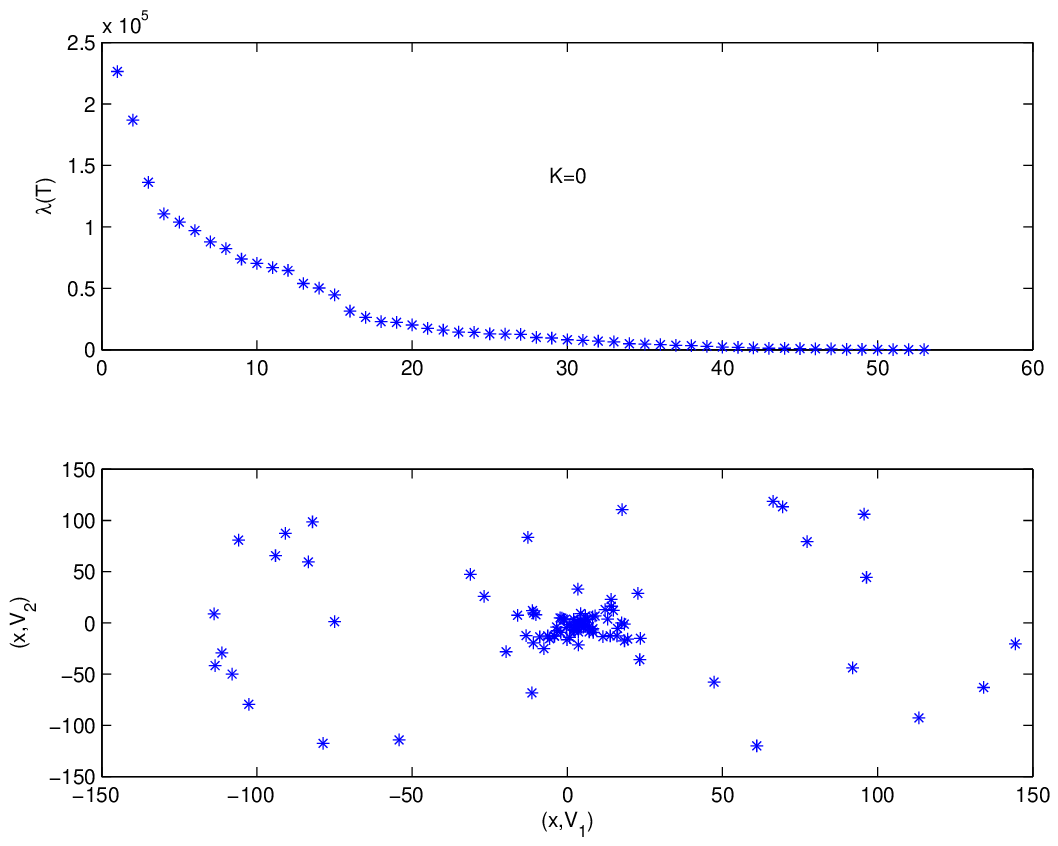}
\caption{Eigenvalues of the matrix $T$ and projection of the dynamics on the
first and second eigenvectors for $K=0$ (triangle interaction)}
\label{geopro_triangle_0}
\end{figure}

In the absence of interaction ($K=0$) the inertial tensor has many large
eigenvalues and the projections of the orbits on the two largest
eigenvectors show no distinctive pattern (Fig. \ref{geopro_triangle_0}).

\begin{figure}[htb]
\centering
\includegraphics[width=0.6\textwidth]{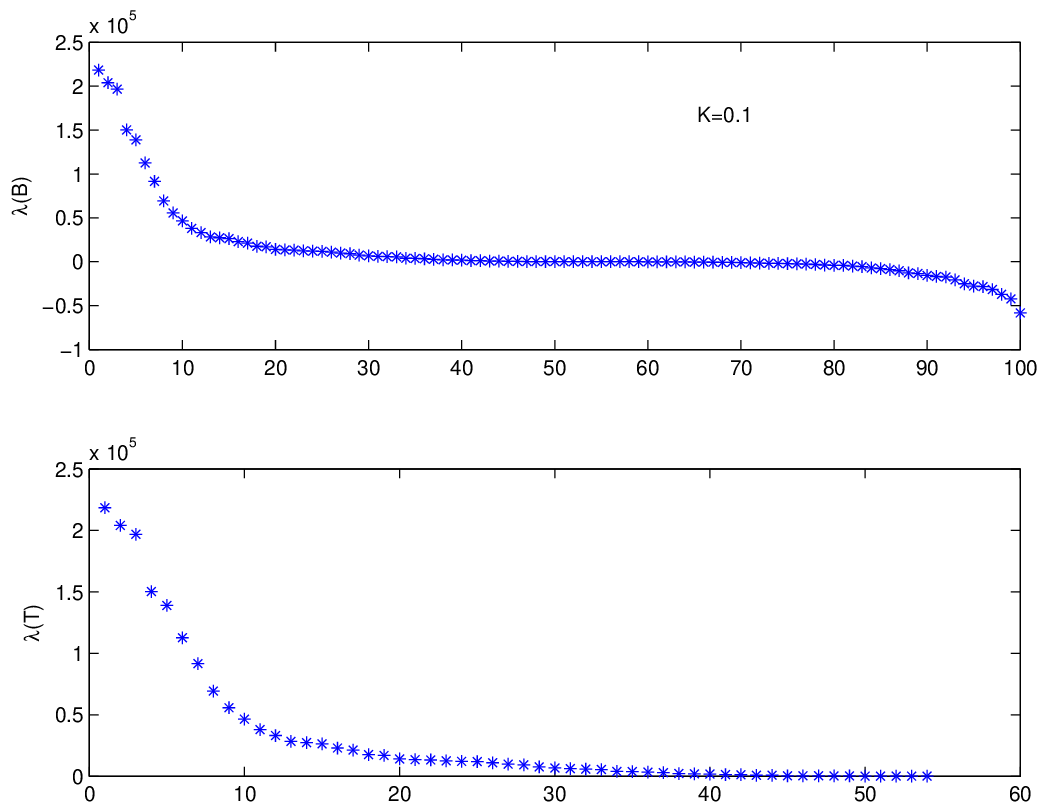}
\caption{Eigenvalues of the $B$ and $T$ matrices for $K=0.1$ (triangle
interaction)}
\label{LambdaBT_triang_0_1}
\end{figure}

One sees that the case $K=0.1$ (Figs. \ref{LambdaBT_triang_0_1} and \ref%
{geopro_triangle_0_1}) is not very different from the $K=0$ case, showing
that strong correlations have not yet developed.

\begin{figure}[htb]
\centering
\includegraphics[width=0.6\textwidth]{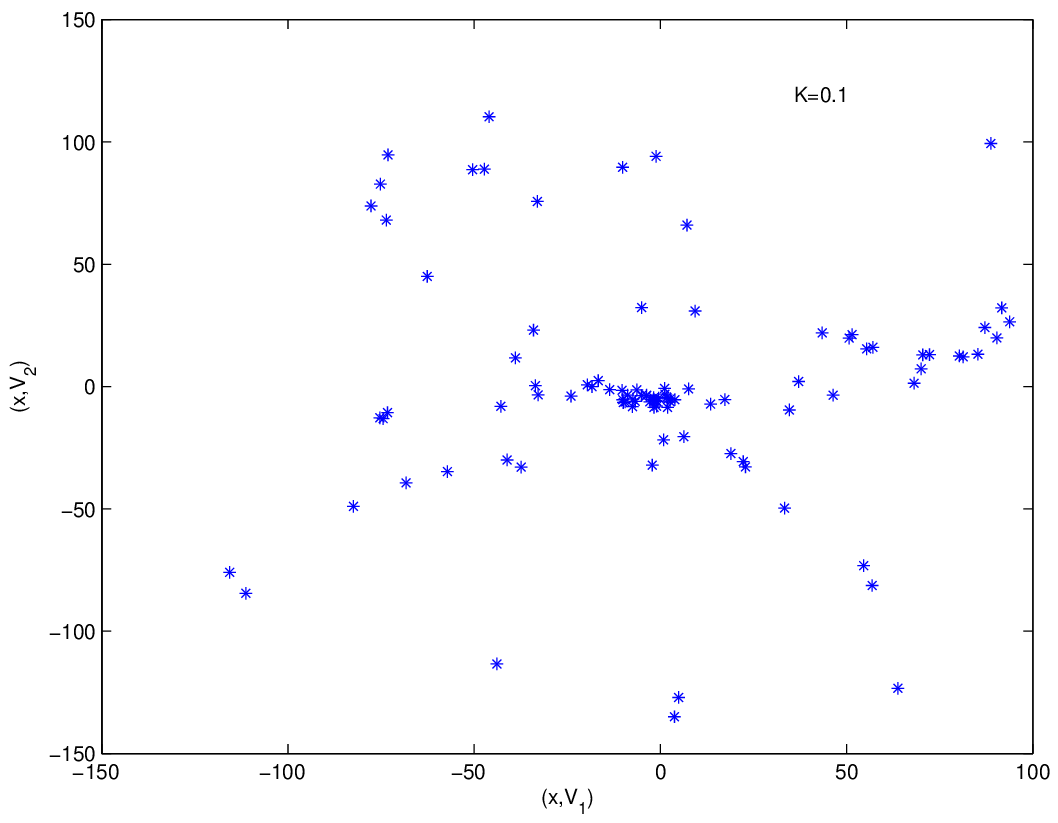}
\caption{Projection of the dynamics on the first and second eigenvectors for 
$K=0.1$ (triangle interaction)}
\label{geopro_triangle_0_1}
\end{figure}
\begin{figure}[htb]
\centering
\includegraphics[width=0.6\textwidth]{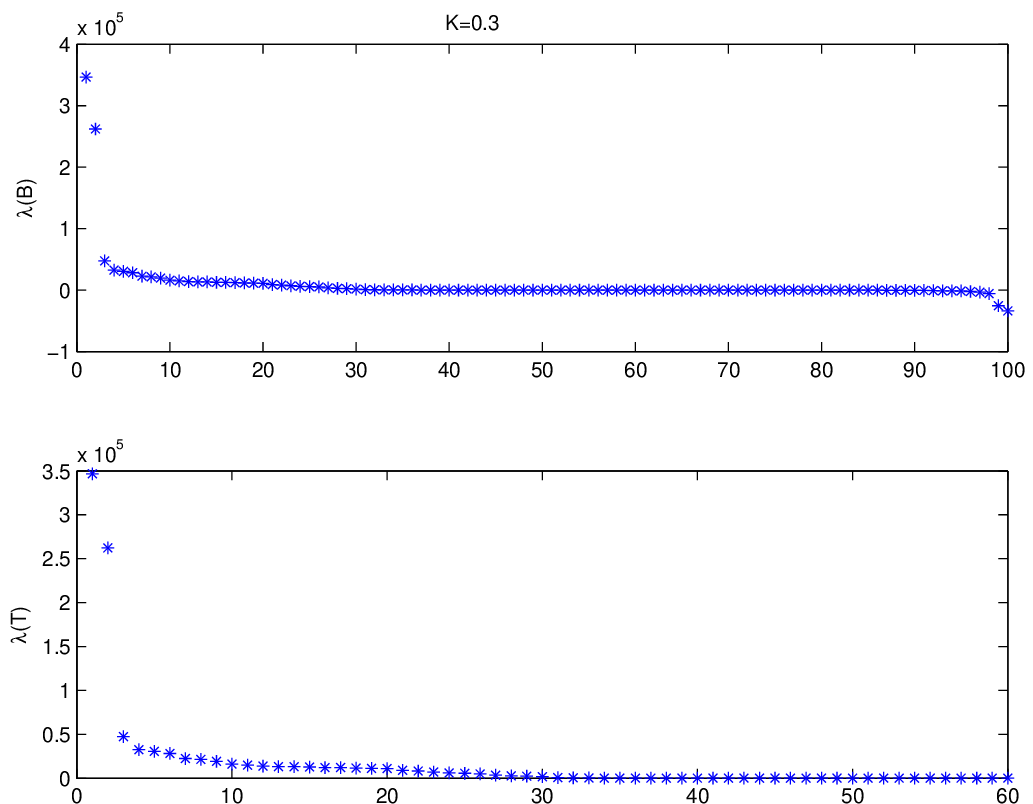}
\caption{Eigenvalues of the $B$ and $T$ matrices for $K=0.3$ (triangle
interaction)}
\label{LambdaBT_triang_0_3}
\end{figure}
\begin{figure}[htb]
\centering
\includegraphics[width=0.6\textwidth]{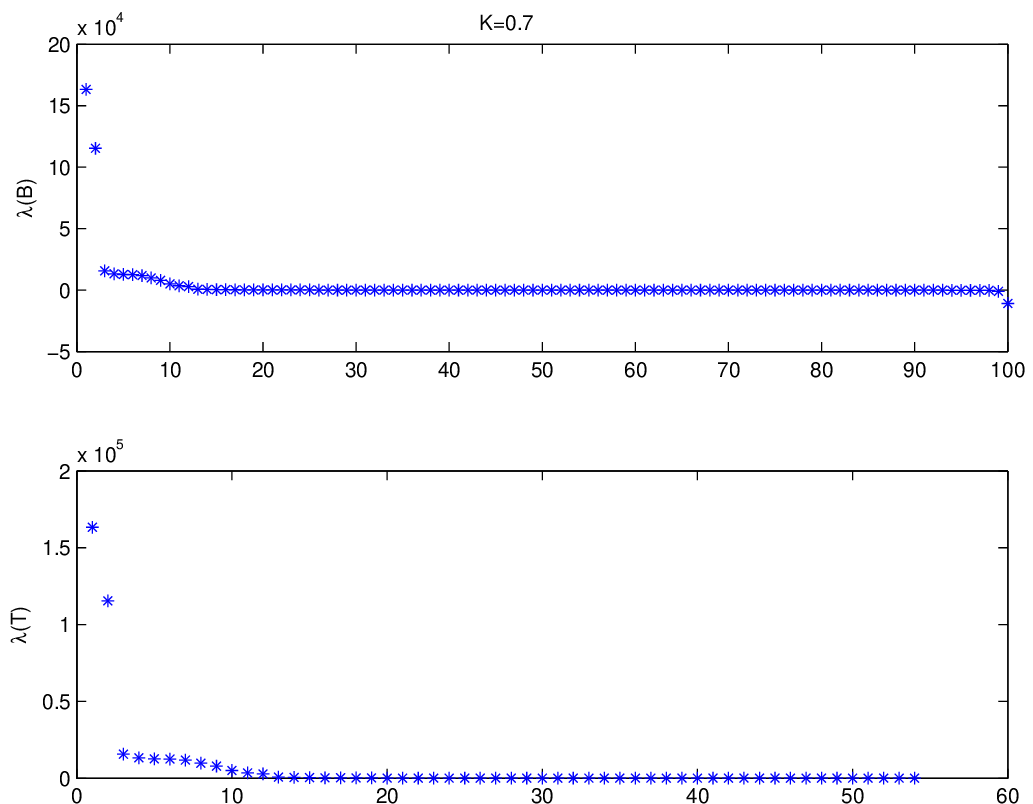}
\caption{Eigenvalues of the $B$ and $T$ matrices for $K=0.7$ (triangle
interaction)}
\label{LambdaBT_triang_0_7}
\end{figure}

It is only for $K=0.3$ and $0.7$ that the dynamics is almost two dimensional
and strongly correlated (Figs. \ref{LambdaBT_triang_0_3}, \ref%
{geopro_triangle_0_3}, \ref{LambdaBT_triang_0_7} and \ref%
{geopro_triangle_0_7}).

\begin{figure}[htb]
\centering
\includegraphics[width=0.6\textwidth]{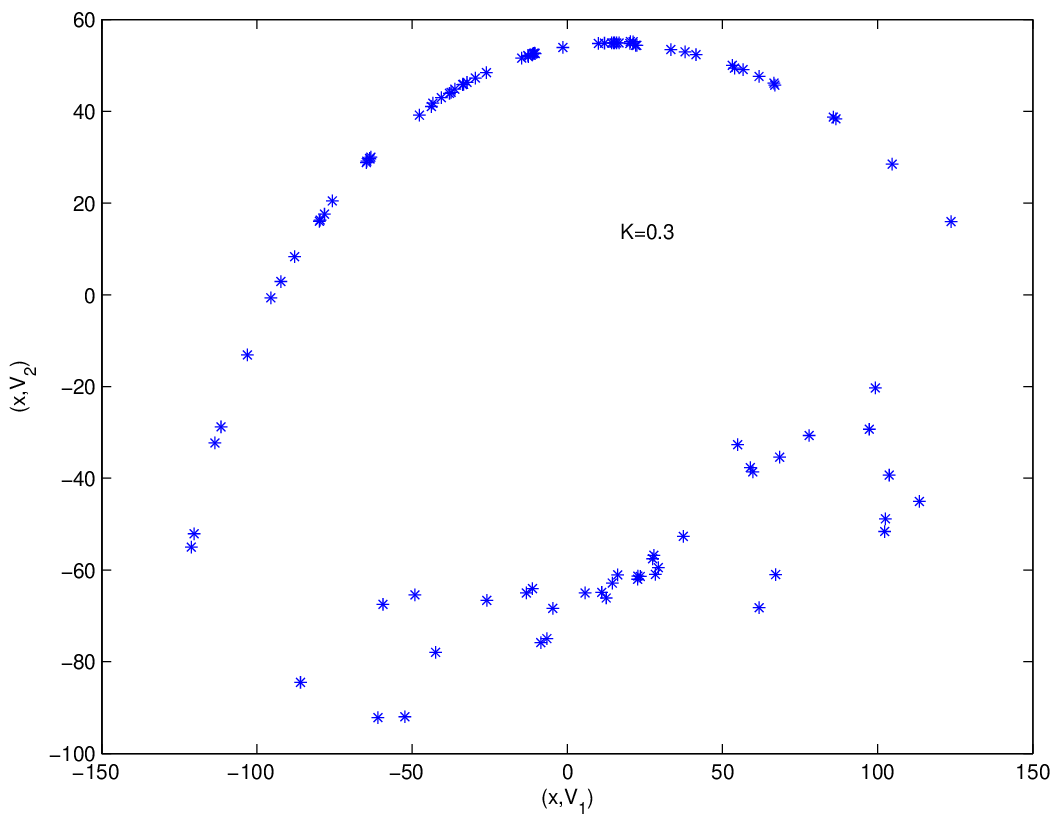}
\caption{Projection of the dynamics on the first and second eigenvectors for 
$K=0.3$ (triangle interaction)}
\label{geopro_triangle_0_3}
\end{figure}

\begin{figure}[htb]
\centering
\includegraphics[width=0.6\textwidth]{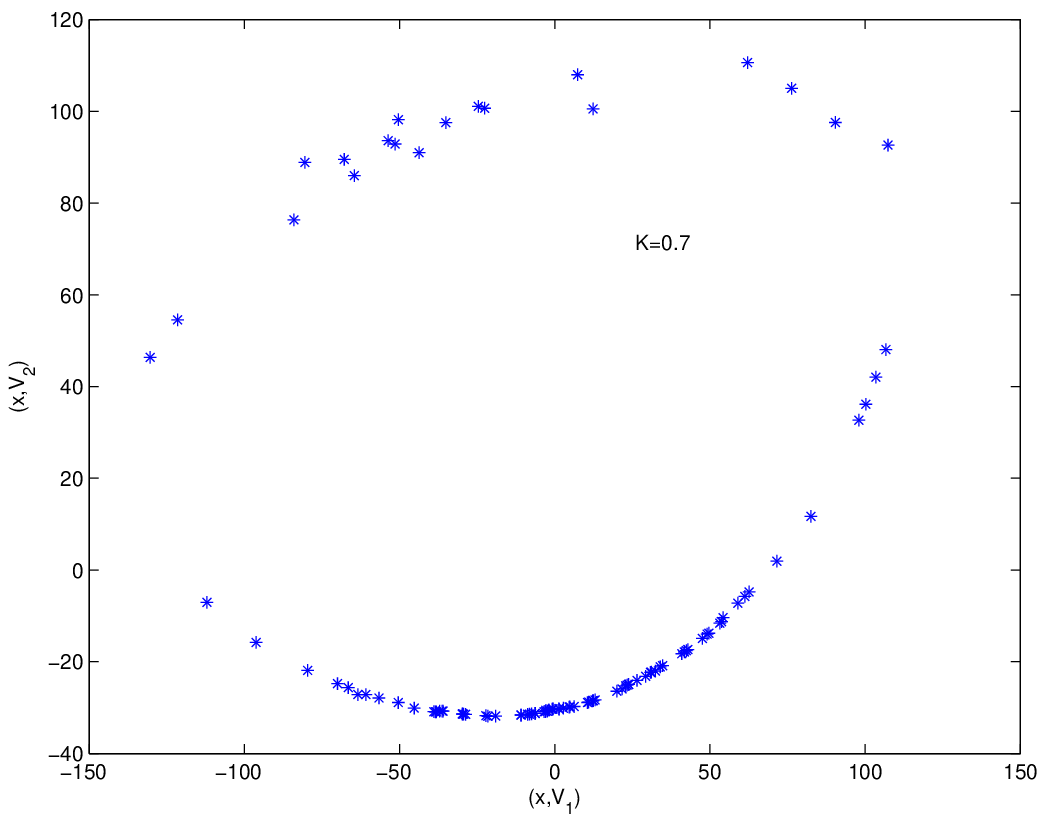}
\caption{Projection of the dynamics on the first and second eigenvectors for 
$K=0.7$ (triangle interaction)}
\label{geopro_triangle_0_7}
\end{figure}

\subsubsection{Dynamical clustering}

\begin{figure}[htb]
\centering
\includegraphics[width=0.6\textwidth]{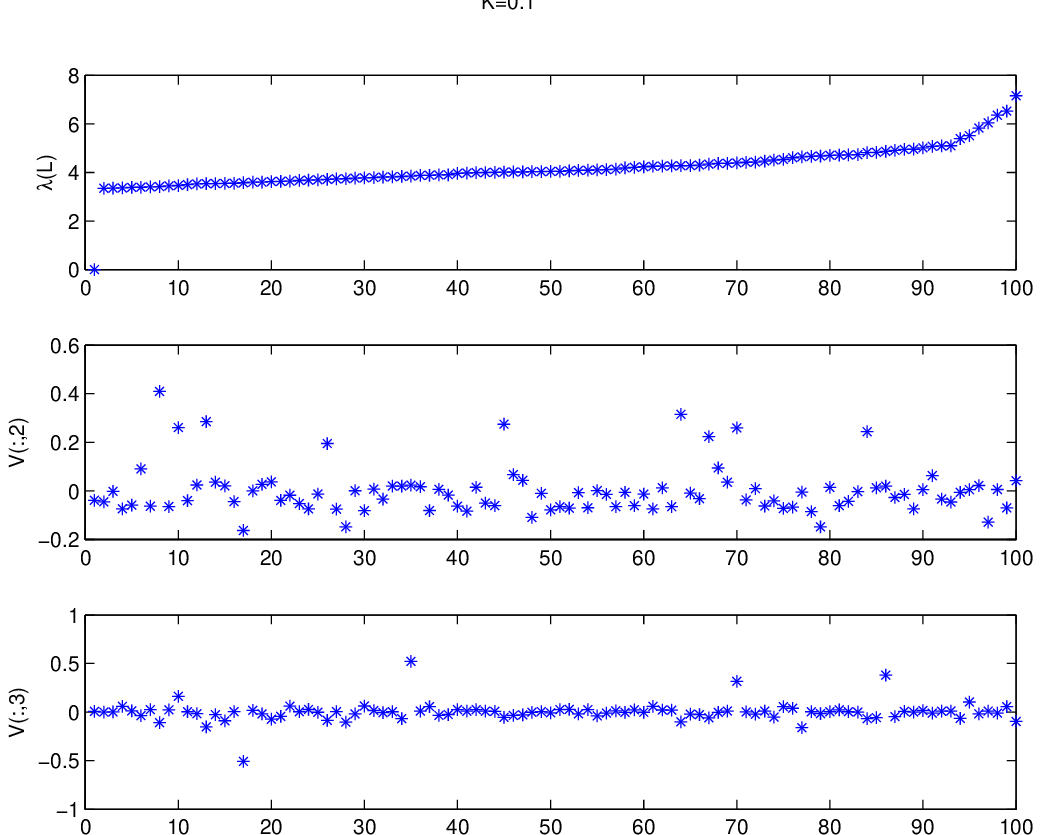}
\caption{The spectrum of the Laplacian matrix $L$ and the second and third
eigenvectors for $K=0.1$ (triangle interaction)}
\label{lspectr_triang_0_1}
\end{figure}

As before the dynamical distances and the adjacency matrix are obtained from
the coordinate increments. The spectrum of the Laplacian matrix $L$ and the
second and third eigenvectors for $K=0.1$, $0.3$ and $0.7$ are displayed in
the Figs. \ref{lspectr_triang_0_1}, \ref{lspectr_triang_0_1} and \ref%
{lspectr_triang_0_1}. Some information is obtained from these results,
mostly for $K=0.3$ and $0.7$, however the analysis of the geometry of the
dynamics performed in the previous subsection seems to be, in this case, a
better way to characterize the correlations.

\begin{figure}[htb]
\centering
\includegraphics[width=0.6\textwidth]{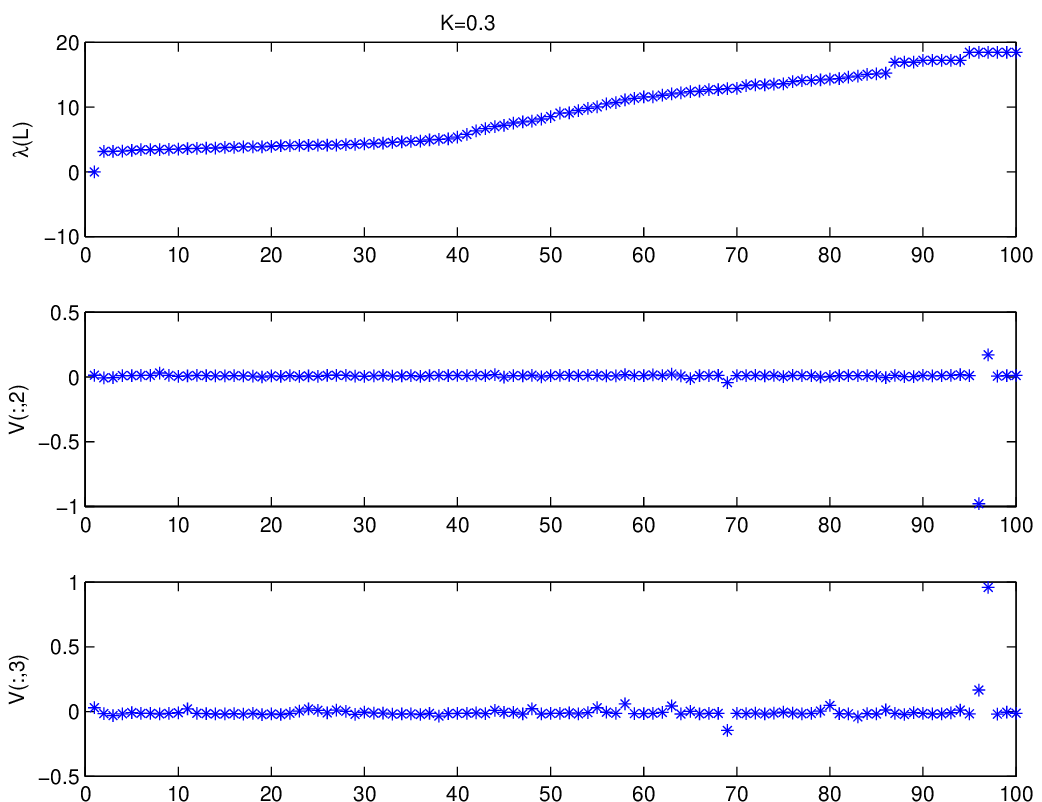}
\caption{The spectrum of the Laplacian matrix $L$ and the second and third
eigenvectors for $K=0.3$ (triangle interaction)}
\label{lspectr_triang_0_3}
\end{figure}

\begin{figure}[htb]
\centering
\includegraphics[width=0.6\textwidth]{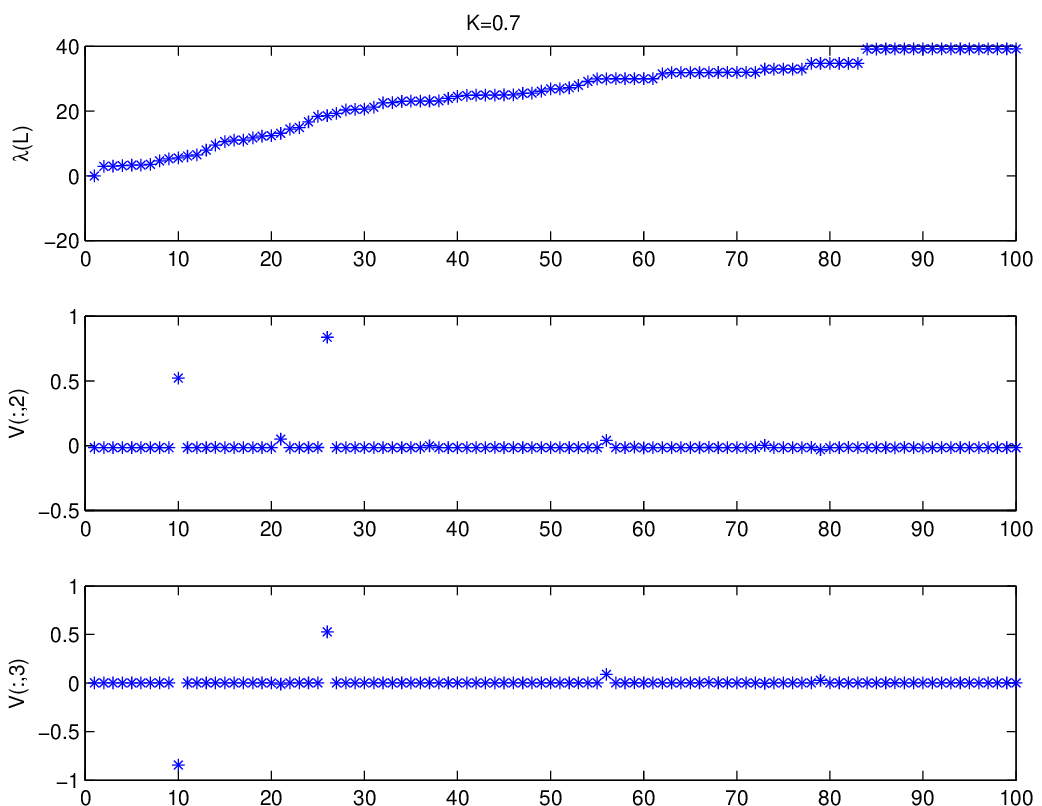}
\caption{The spectrum of the Laplacian matrix $L$ and the second and third
eigenvectors for $K=0.7$ (triangle interaction)}
\label{lspectr_triang_0_7}
\end{figure}
\clearpage

\subsection{The deterministic integrate-and-fire model}

The deterministic integrate-and-fire model is of a different nature as
compared to the two previous models. It suggests that in addition to the
geometric and clustering methods, that are fairly successful for continuous
variable models, other tools should be developed to handle pulsing systems
of this type.

\begin{figure}[htb]
\centering
\includegraphics[width=0.6\textwidth]{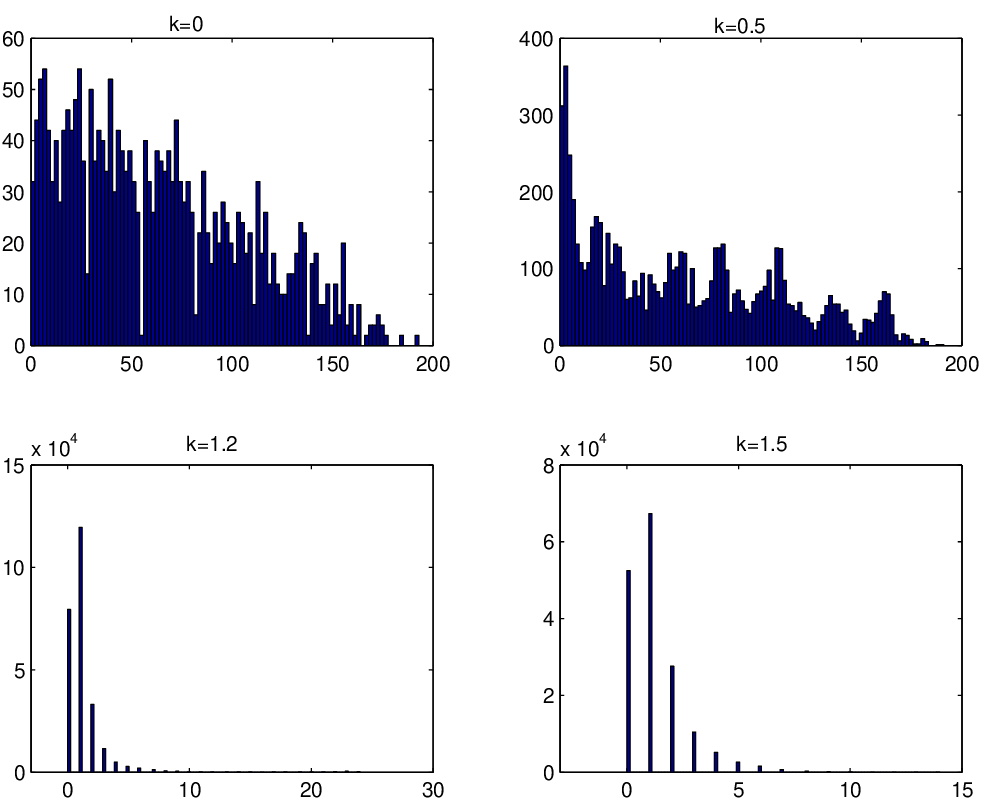}
\caption{Histograms of the firing delays}
\label{delays_inte_fire}
\end{figure}

To put into evidence the firing patterns we have displayed the histograms of
the firing delays (Fig. \ref{delays_inte_fire}). Here we define the firing
delays as the separation in time of each firing with the closest firing in
any one of the other units.

One sees a tendency to organization of the system on the concentration of
the distribution towards smaller delays on passing from $k=0$ to $k=0.5$,
but it only only after $k\approx 1$ that the firings organize into a set of
well defined patterns.

\begin{figure}[htb]
\centering
\includegraphics[width=0.6\textwidth]{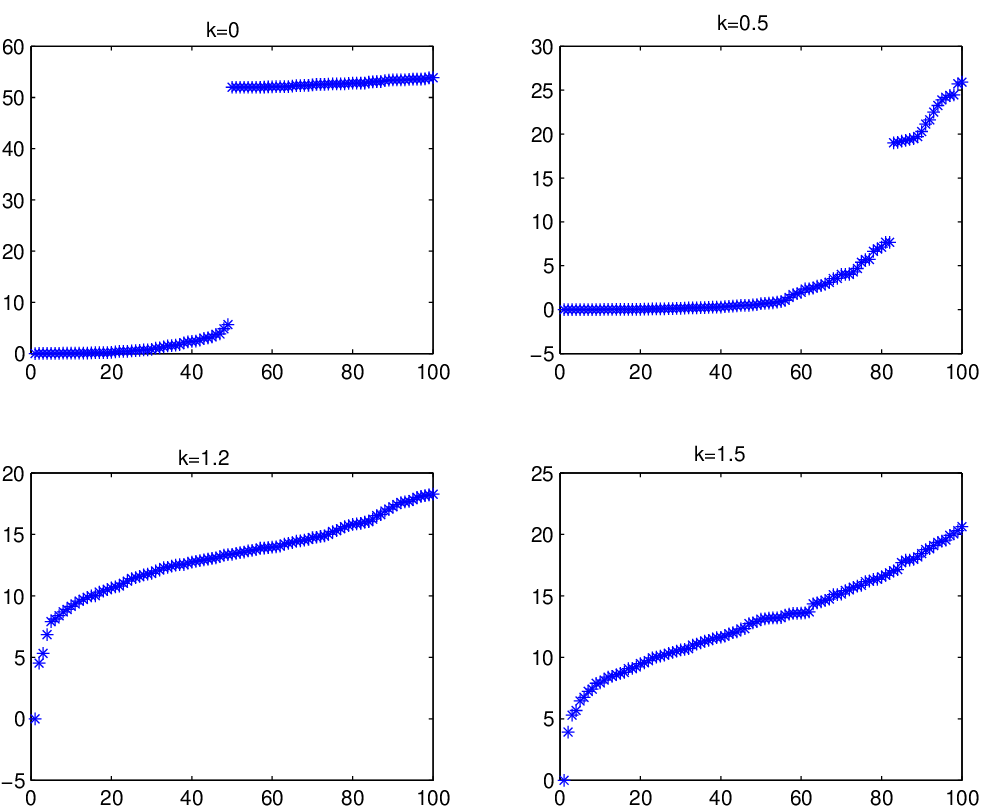}
\caption{The Laplacian matrix spectrum}
\label{clusters_inte_fire}
\end{figure}

From the firing delays a distance between the units may be defined by the
mean of the delays between each pair of units. From the distances an
adjacency matrix was constructed and the spectrum of the Laplacian matrix
computed (Fig. \ref{clusters_inte_fire}). Some information on the firing
clusters is indeed obtained for $k=1.2$ and $1.5$, however the information
provided by the histograms of the firing delays is sharper.

\clearpage

\section{Conclusion}

From the models analyzed in this paper it is clear that, in addition to
synchronization, other types of strongly correlated behavior emerge in the
collective dynamics of interacting systems. What at times has been dismissed
as incoherent behavior contains important collective phenomena that enslave
the dynamics. Hence, it seems important to develop tools which might be able
to characterize qualitative and quantitatively the collective correlation
effects that emerge before or instead of synchronization. A first step in
this direction has been taken in this paper using geometrical and ergodic
techniques.


\begin{thebibliography}{99}
\bibitem{Pikov1} A. Pikovsky, M. Rosenblum, J. Kurths; \textit{%
Synchronization: a universal concept in nonlinear sciences, }Cambridge
University Press 2001.

\bibitem{Pikov2} A. Pikovsky and M. Rosenblum; \textit{Dynamics of globally
coupled oscillators: Progress and perspectives}, Chaos 25 (2015) 097616.

\bibitem{Wu} C. W. Wu; \textit{Synchronization in Complex Networks of
Nonlinear Dynamical Systems, }World Scientific, Singapore 2007.

\bibitem{Kocarev} L. Kocarev (Ed.); \textit{Consensus and Synchronization in
Complex Networks}, Springer, Berlin 2013.

\bibitem{Luo} A. C. J. Luo; \textit{Dynamical System Synchronization},
Springer 2013.

\bibitem{Aguirre} C. Aguirre, D. Campos, P. Pascual and E. Serrano; \textit{%
Synchronization effects using a piecewise linear map-based spiking bursting
neuron model}, Neurocomputing 69 (2006) 1116-1119.

\bibitem{Koronov} A. A. Koronovskii,O. I. Moskalenko, V. I. Ponomarenko, M.
D. Prokhorov, A. E. Hramov; \textit{Binary generalized synchronization},
Chaos, Solitons and Fractals 83 (2016) 133-139.

\bibitem{Glass} L. Glass, \textit{Synchronization and rhytmic processes in
physiology}, Nature 410 (2001) 277--284.

\bibitem{Hayes} D. L. Hayes, S. J. Asirvatham and P. A. Friedman; \textit{%
Cardiac Pacing, Defibrillation and Resynchronization: A Clinical Approach, }%
Wiley-Blackwell, Chichester 2013.

\bibitem{circadian} Z. Lu, K. Klein-Carde\~{n}a, S. Lee, T. M. Antonsen, M.
Girvan and E. Ott; \textit{Resynchronization of circadian oscillators and
the east-west asymmetry of jet-lag}, Chaos 26 (2016) 094811.

\bibitem{Schnitzler} A. Schnitzler and J. Gross; \textit{Normal and
pathological oscillatory communication in the brain}, Nature Reviews -
Neuroscience 6 (2005) 285-295.

\bibitem{Ortega} G. J. Ortega, L. M. de la Prida, R. G. Sola and J. Pastor; 
\textit{Synchronization clusters of interictal activity in the lateral
temporal cortex of epileptic patients: Intraoperative electrocorticographic
analysis}, Epilepsia 49 (2008) 269-280.

\bibitem{Ioannides} A. A. Ioannides, V. Poghosyan, J. Dammers and M. Streit; 
\textit{Real-time neural activity and connectivity in healthy individuals
and schizophrenia patients}, NeuroImage 23 (2004) 473-482.

\bibitem{Sawa} A. Sawa and S. H. Snyder; \textit{Schizophrenia: Diverse
approaches to a complex disease}, Science 296 (2002) 692-695.

\bibitem{Roelfsema} P. R. Roelfsema, A. K. Engel, P. K\"{o}nig and W.
Singer; \textit{Visuomotor integration is associated with zero time-lag
synchronization among cortical areas}, Nature 385 (1997) 157-161.

\bibitem{Lachaux} J.-P. Lachaux, E. Rodriguez, J. Martinerie and F. J.
Varela; \textit{Measuring phase synchrony in brain signals}, Human Brain
Mapping 8 (1999) 194-208.

\bibitem{Long} M. A. Long, C. E. Landisman and B. W. Connors; \textit{Small
Clusters of Electrically Coupled Neurons Generate Synchronous Rhythms in the
Thalamic Reticular Nucleus}, The Journal of Neuroscience 24 (2004) 341-349.

\bibitem{Buck} J. Buck; \textit{Synchronous rhythmic flashing of fireflies},
The Quartely Review of Biology 63 (1988) 265-289.

\bibitem{Varela1} E. Rodriguez, N. George, J.-P. Lachaux, J. Martinerie, B.
Renault and F. Varela; \textit{Perception's shadow: long-distance
synchronization of human brain activity}, Nature 397 (1999) 430-433.

\bibitem{Varela2} F. Varela, J.-P. Lachaux, E. Rodriguez and J. Martinerie, 
\textit{The brainweb: phase synchronization and large-scale integration},
Nature Reviews Neuroscience 2 (2001) 229-239.

\bibitem{Engel} A. K. Engel and W. Singer; \textit{Temporal binding and the
neural correlates of sensory awareness}, Trends in Cognitive Sciences 5
(2001) 16-25.

\bibitem{computation} D. Malagarriga, M. A. Garc\'{\i}a-Vellisca, A. E. P.
Villa, J. M. Buld\'{u}, J. Garc\'{\i}a-Ojalvo and A. J. Pons; \textit{%
Synchronization-based computation through networks of coupled oscillators},
Frontiers in Computational Neuroscience 9:97, 2015.

\bibitem{quantum} G. Manzano, F. Galve, G. L. Giorgi, E. Hern\'{a}ndez-Garc%
\'{\i}a and R. Zambrini; \textit{Synchronization, quantum correlations and
entanglement in oscillator networks}, Scientific Reports 3 : 1439 DOI:
10.1038/srep01439 1.

\bibitem{synchro1} R. Sevilla-Escoboza, J. M. Buld\'{u}, S. Boccaletti, D.
Papo, D.-U. Hwang, G. Huerta-Cuellar and R. Guti\'{e}rrez; \textit{%
Experimental implementation of maximally synchronizable networks},
arXiv:1507.02551.

\bibitem{synchro2} A. Navas, J. A. Villacorta-Atienza, I. Leyva, J. A.
Almendral, I. Sendi\~{n}a-Nadal and S. Boccaletti; \textit{Effective
centrality and explosive synchronization in complex networks}, Phys. Rev. E
92 (2015) 062820.

\bibitem{synchro3} M. Golubitsky and I. Stewart; \textit{Rigid patterns of
synchrony for equilibria and periodic cycles in network dynamics}, Chaos 26
(2016) 094803.

\bibitem{synchro4} B. Ottino-L\"{o}ffler and S. H. Strogatz; \textit{%
Frequency spirals}, Chaos 26 (2016) 094804.

\bibitem{synchro5} P. S. Skardal, D. Taylor and J. Sun; \textit{Optimal
synchronization of directed complex networks}, Chaos 26 (2016) 094807.

\bibitem{synchro6} L. Wang and G. Chen; \textit{Synchronization of
multi-agent systems with metric-topological interactions}, Chaos 26 (2016)
094809.

\bibitem{synchro7} J. Emenheiser, A. Chapman, M. P\'{o}sfai, J. P.
Crutchfield, M. Mesbahi and R. M. D'Souza; \textit{Patterns of patterns of
synchronization: Noise induced attractor switching in rings of coupled
nonlinear oscillators}, Chaos 26 (2016) 094816.

\bibitem{layer1} V. V. Makarova, A. A. Koronovskii, V. A. Maksimenko, A. E.
Hramov, O. \ I. Moskalenko, J. M. Buld\'{u} and S. Boccaletti; \textit{%
Emergence of a multilayer structure in adaptive networks of phase oscillators%
}, Chaos, Solitons and Fractals 84 (2016) 23-30.

\bibitem{layer2} R. Sevilla-Escoboza, I. Sendi\~{n}a-Nadal, I. Leyva, R. Guti%
\'{e}rrez, J.M. Buld\'{u} and S. Boccaletti; \textit{Inter-layer
synchronization in multiplex networks}, arXiv: 1510.07498.

\bibitem{Fujisaka} H. Fujisaka and T. Yamada; \textit{Stability theory of
synchronized motion in coupled-oscillator systems}, Prog. of Theor. Phys. 69
(1983) 32-47.

\bibitem{Pecora1} L. M. Pecora and T. L. Carroll; \textit{Synchronization in
chaotic systems}, Physical Review Letters 64 (1990) 821--824.

\bibitem{Vilela1} R. Vilela Mendes; \textit{Clustering and synchronization
with positive Lyapunov exponents}, Physics Letters A257 (1999) 132-138.

\bibitem{Bocca} S. Boccaletti, J. Kurths, G. Osipov, D. L. Valladares and C.
S. Zhou; \textit{The synchronization of chaotic systems}, Physics Reports
366 (2002) 1--101.

\bibitem{Pecora2} L. M. Pecora and T. L. Carroll; \textit{Synchronization of
chaotic systems}, Chaos 25 (2015) 097611.

\bibitem{mobile} N. Fujiwara, J. Kurths and A. D\'{\i}az-Guilera; \textit{%
Synchronization of mobile chaotic oscillator networks}, Chaos 26 (2016)
094824.

\bibitem{Rodriguez} M. Lopez and F. Rodriguez; \textit{Detection Method for
Phase Synchronization in a Population of Spiking Neurons},\ IWINAC 2013,
Part I, LNCS 7930, J.M. Ferr\'{a}ndez et al. (Eds.), pp. 421--431,
Springer-Verlag Berlin Heidelberg 2013..

\bibitem{chimera1} Y. Kuramoto and D. Battogtokh; \textit{Coexistence of
Coherence and Incoherence in Nonlocally Coupled Phase Oscillators},
Nonlinear Phenom. Complex Syst. 5 (2002) 380-385.

\bibitem{chimera2} F. P. Kemeth, S. W. Haugland, L. Schmidt, I. G.
Kevrekidis and K. Krischer; \textit{A classification scheme for chimera
states}, Chaos 26 (2016) 094815.

\bibitem{chimera3} J. D. Hart, K. Bansal, T. E. Murphy and R. Roy; \textit{%
Experimental observation of chimera and cluster states in a minimal globally
coupled network}, Chaos 26 (2016) 094801.

\bibitem{chimera4} D. M. Abrams and S. H. Strogatz; \textit{Chimera States
for Coupled Oscillators}, Phys. Rev. Lett. 93 (2004) 174102.

\bibitem{chimera5} E. A. Martens, C. Bick and M. J. Panaggio; \textit{%
Chimera states in two populations with heterogeneous phase-lag}, Chaos 26
(2016) 094819.

\bibitem{chimera6} S. Nkomo, M. R. Tinsley and K. Showalter; \textit{Chimera
and chimera-like states in populations of nonlocally coupled homogeneous and
heterogeneous chemical oscillators}, Chaos 26 (2016) 094826.

\bibitem{cluster1} I. Franovic, K. Todorovic, N. Vasovic and N. Buric; 
\textit{Cluster synchronization of spiking induced by noise and interaction
delays in homogenous neuronal ensembles}, Chaos 22 (2012) 033147.

\bibitem{cluster2} S. Jalan, A. Kumar, A. Zaikin and J. Kurths; \textit{%
Interplay of degree correlations and cluster synchronization}, Phys. Rev. E
94 (2016) 062202.

\bibitem{cluster3} M. T. Schaub, N. O'Clery, Y. N. Billeh, J.-C. Delvenne,
R. Lambiotte and M. Barahona; \textit{Graph partitions and cluster
synchronization in networks of oscillators}, Chaos 26 (2016) 094821.

\bibitem{cluster4} F. Sorrentino and L. Pecora; \textit{Approximate cluster
synchronization in networks with symmetries and parameter mismatches}, Chaos
26 (2016) 094823.

\bibitem{cluster5} T. Nishikawa and A. E. Motter; \textit{Network-complement
transitions, symmetries, and cluster synchronization}, Chaos 26 (2016)
094818.

\bibitem{cluster6} P. Ji, T. K. DM. Peron, F. A. Rodrigues and J. Kurths; 
\textit{Analysis of cluster explosive synchronization in complex networks},
arXiv:1402.5587.

\bibitem{WataStro} S. Watanabe and S. H. Strogatz, \textit{Integrability of
a Globally Coupled Oscillator Array}, Phys. Rev. Lett. 70 (1993) 2391-2394.

\bibitem{Antonsen} E. Ott and T. M. Antonsen; \textit{Low dimensional
behavior of large systems of globally coupled oscillators}, Chaos 18 (2008)
037113.

\bibitem{Kuramoto} Y. Kuramoto; \textit{Chemical Oscillations, Waves, and
Turbulence}, Springer (1984)

\bibitem{Strogatz1} S. H. Strogatz; \textit{From Kuramoto to Crawford:
exploring the onset of synchronization in populations of coupled oscillators}%
, Physica D 143 (2000) 1--20.

\bibitem{Acebron} J. Acebr\'{o}n, L. L. Bonilla, C. J. P\'{e}rez Vicente, F.
Ritort and R. Spigler; \textit{The Kuramoto model: A simple paradigm for
synchronization phenomena}, Review of Modern Physics 77 (2005) 137--185

\bibitem{Vilela3} R. Vilela Mendes; \textit{Tools for network dynamics}
International Journal of Bifurcation and Chaos 15 (2005) 1185-1213.

\bibitem{Restrepo} J. G. Restrepo, E. Ott and B. R. Hunt; \textit{Onset of
synchronization in large networks of coupled oscillators}, Physical Review E
71 (2005) 036151.

\bibitem{VilelaChaos} R. Vilela Mendes; \textit{Ergodic parameters and
dynamical complexit}y, Chaos 21 (2011) 037115.

\bibitem{vonLuxburg} U. Von Luxburg; \textit{A tutorial on spectral
clustering}, Stat. Comput. 17 (2007) 395-416.

\bibitem{VilelaCondExp} R. Vilela Mendes; \textit{Conditional exponents,
entropies and a measure of dynamical self-organization}, Physics Letters A
248 (1998) 167.

\bibitem{VilelaSelfOrg} R. Vilela Mendes; \textit{Characterizing
self-organization and coevolution by ergodic invariants}, Physica A 276
(2000) 550.
\end{thebibliography}
\end{document}